\documentclass[usenatbib]{mn2e} 
\usepackage{epsfig}
\newcommand{\lsim}{\lower 2pt \hbox{$\, \buildrel {\scriptstyle<}\over {\scriptstyle \sim}\,$}}  
\newcommand{\gsim}{\lower 2pt\hbox{$\, \buildrel {\scriptstyle >}\over {\scriptstyle \sim}\,$}}
\defcitealias{Veilleux99a}{Veilleux et al., 1999}
\defcitealias{Sanders88a}{Sanders et al. (1988a)}
\defcitealias{Sanders88b}{Sanders et al. (1988b)}
\defcitealias{Papovich06}{Papovich et al. (2006)}
\defcitealias{Tacconi08}{Tacconi et al., 2008}
\defcitealias{Foster04}{F{\"o}rster Schreiber et al., 2004}
\bibpunct{(}{)}{;}{a}{,}{,} \title[Optical spectroscopy of ULIRGs: 
star formation histories and evolution.]{The properties of the stellar
populations in ULIRGs II: star formation histories and evolution}

\author[J.Rodr\'iguez Zaur\'in, C.N.Tadhunter and R.M.Gonz\'alez
Delgado$^{2}$]{J.Rodr\'iguez
Zaur\'in$^{1,2}$\thanks{E-mail:jrz@damir.iem.csic.es}, C.N
Tadhunter$^{2}$ and R.M.Gonz\'alez Delgado$^{3}$\\ $^{1}$Department of
molecular and infrared astronomy, IEM (CSIC), 28006 Madrid, SPAIN\\
$^{2}$ Department of physics and Astronomy, University of Sheffield, Sheffield S3 7RH\\
$^{3}$Instituto de Astrofisica de Andalucia(CSIC), P.O.Box 3004, 18080
Granada, Spain}

\begin{document}

\pagerange{\pageref{firstpage}--\pageref{lastpage}} \pubyear{2002}

\maketitle

\label{firstpage}

\begin{abstract}

This is the second of two papers presenting a detailed long-slit
spectroscopic study of the stellar populations in a sample of 36
ULIRGs. In the previous paper we presented the sample, the data and
the spectral synthesis modelling. In this paper we carry out a more
detailed analysis of the modelling results, with the aim of
investigating the general properties of the stellar populations
(i.e. age, reddening and percentage contribution) and the evolution of
the host galaxies, comparing the results with other studies of ULIRGs
and star forming galaxies in the high-z Universe. The characteristic
age of the young stellar populations (YSPs) is $\leq$ 100 Myr in the
nuclei of the overwhelming majority of galaxies, consistent with the
characteristic timescale of the major burst of star formation
associated with the final stages of major galaxy mergers. However, the
modelling results clearly reveal that the star formation histories of
ULIRGs are complex, with at least two epochs of star formation
activity. Overall, these results are consistent with models that
predict an epoch of enhanced star formation coinciding with the first
pass of the merging nuclei, along with a further, more intense,
episode of star formation occurring as the nuclei finally merge
together. It is also found that, although YSPs make a major
contribution to the optical emission in most of the extended and
nuclear apertures examined, they tend to be younger and more reddened
in the nuclear regions of the galaxies. This is in good agreement with
the merger simulations, which predict that the bulk of the star
formation activity in the final stages of mergers will occur in the
nuclear regions of the merging galaxies. In addition, our results show
that ULIRGs have total stellar masses that are similar to, or smaller
than, the break of the galaxy mass function (i.e. ULIRGs are
sub-$m_{\ast}$ or $\sim m_{\ast}$ systems), and that the young stellar
populations detected at optical wavelengths dominate the stellar mass
contents of the galaxies.  Finally, we find no significant differences
between the ages of the YSP in ULIRGs with and without optically
detected Seyfert nuclei, nor between those with warm and cool mid- to
far-IR colours.  While this results do not entirely rule out the idea
that cool ULIRGs with HII/LINER spectra evolve into warm ULIRGs with
Seyfert-like spectra, it is clear that the AGN activity in local
Seyfert-like ULIRGs has not been triggered a substantial period
($\geq$100~Myr) {\it after} the major merger-induced starbursts in the
nuclear regions.

\end{abstract}

\begin{keywords}
Galaxies: evolution -- galaxies: starburst.
\end{keywords}

\section{Introduction}

First discovered in large numbers by the {\it IRAS} satellite,
ultraluminous infrared galaxies (ULIRGs, L$_{\rm IR} >$ 10$^{12}$
L$_{\odot}$) represent some of the most rapidly evolving objects in
the local Universe. These objects are important in several respects,
such as testing the merger models, investigating the nature of the
AGN-starburst connection in merging systems, and understanding the
physical processes taking place in the recently discovered
high-redshift star forming galaxies, which have similar properties to
the nearby ULIRGs.
  
 A remarkable property of ULIRGs is that more than 90\% of them are
associated with galaxy mergers and interactions \cite[see][for a
review]{Sanders96}. The simulations have shown that, during galaxy
collisions involving gas-rich disk galaxies, the gas loses angular
momentum due to dynamical friction, tidal torques, and inflows towards
the centres of the galaxies \cite[e.g.][]{Mihos96}. Such high nuclear
concentrations of gas are capable of triggering both starburst and AGN
activity. In general, these simulations predict two epochs of enhanced
star formation in major galaxy mergers: the first occurring around the
first encounter, and the second episode when the nuclei are close
to coalescence \citep{Mihos96,Barnes96,Springel05,Cox06,PDiMatteo07}
. However, both the time lag and the relative intensity of the peaks
of starburst activity during the merger event depend on several
factors, such as the presence of bulges, feedback effects, gas content
and orbital geometry. Given that the models make specific predictions
about the histories of the star formation triggered in the course of
major gas-rich mergers, studies of the stellar populations in ULIRGs
provide useful information about the mergers and, potentially, allow
us to test the models.

ULIRGs are classified on the basis of their infrared colours as warm
({\it f\/}$_{25}$/{\it f}$_{60} >$ 0.2)\footnote{ The quantities {\it
f\/}$_{25}$ and {\it f}$_{60}$ represent the {\it IRAS} flux densities
in Janskys at 25$\mu$m and 60$\mu$m.} and cool ({\it f\/}$_{25}$/{\it
f}$_{60} \leq$ 0.2) ULIRGs. Warm ULIRGs represent $\sim$20 -- 25\% of
the total population of ULIRGs discovered by {\it IRAS}. Most of such
objects have an AGN optical spectral classfication, tend to be more
compact than cool ULIRGs, and are frequently found in an advanced
merger state \citep{Surace98b}. These properties suggest that ULIRGs
play an important role in the formation and evolution of QSOs (Sanders
et al., 1988b) and radio galaxies \citep{Tadhunter05}. Indeed,
\citetalias{Sanders88b} proposed a scenario in which cool ULIRGs
evolve into warm ULIRGs on their way to becoming typical
optically-selected QSOs. Further evidence in favour of such an
evolutionary scenario was reported by \cite{Surace98b},
\cite{Surace99}, \cite{Surace00a} and \cite{Surace00b} on the basis of
their imaging studies of cool and warm ULIRGs, and
\cite{Canalizo00a,Canalizo00b,Canalizo01}, based on a spectroscopic
analysis of a sample of transition QSOs --- objects that, based on
their infrared colors, may represent a transitionary stage between
ULIRGs and QSOs. More recently, \cite{Tadhunter07} carried out a mid-
to far-IR study of powerful radio galaxies, searching for signs of
hidden star formation activity. They found that the fraction of
powerful radio galaxies with energetically significant star formation
activity (20\% -- 30\%) is similar to that deduced from optical
studies. The derived ages and masses of the young stellar populations
in the \cite{Wills04}, \cite{Tadhunter05} and \cite{Tadhunter07}
samples of radio galaxies, and the fact that some of these radio
galaxies are actually classified as ULIRGs, is consistent with the
idea that some radio galaxies are the evolved product of ULIRGs
\citep{Tadhunter05}.

In terms of the end products of the merger events associated with ULIRGs, 
the kinematic studies of \cite{Genzel01} and \cite{Tacconi02}
have shown that ULIRGs match the locations of intermediate mass, disky
ellipticals and S0 galaxies in the fundamental plane. ULIRGs have
effective radii and velocity dispersions smaller than those of their
comparison sample of nearby radio galaxies. Furthermore, the
dynamical masses of ULIRGs are 10$^{10}$ -- 10$^{11}$ M$_{\odot}$
\citep{Tacconi02,Colina05,Dasyra06a,Dasyra06b}, which is similar to
intermediate-mass elliptical galaxies. Based on the dynamical
properties of ULIRGs, \cite{Tacconi02} concluded that not all ULIRGs
can evolve into typical QSOs or powerful radio galaxies. However, the
recent work of \cite{Dasyra07}, based on near-IR spectroscopic
observations of a sample of 12 (mainly Palomar Green) QSOs, has shown
that it is possible for ULIRGs to have similar dynamical properties to
QSOs. Therefore, it is clear that, despite the recognition of an
evolutionary link between ULIRGs, QSOs/radio galaxies and elliptical
galaxies, considerable uncertainties remain about the true nature of
the link.

Studies of ULIRGs are also important in the context of the various
populations of high-z galaxies that have been recently detected using
near-IR \cite[distant red galaxies, DRGs, ][]{Franx03}, mid-IR
\cite[24${\mu}$m {\it Spitzer}-selected
galaxies, ][and others]{Caputi06a,Yan07,Sajina07},
and sub-millimeter wavelengths \cite[sub-mm galaxies, SMGs, ][and
others]{Smail02,Blain02,Chapman03,Pope05}. These objects have
properties similar to the nearby ULIRGs (stellar mass and luminosity,
for example), and make up a substantial fraction of the star formation
(SF) at z $\sim$ 1-2. As local analogues of such objects, ULIRGs
provide an opportunity to study the physical processes associated with
their prodigious SF activity in depth.

It is also important to emphasize that most of the studies of the
stellar populations in high redshift galaxies are based on the
modelling of a few photometric points and, therefore, it is hard to
constrain the properties of these populations. Sensitive studies of
star forming galaxies at lower redshifts are useful, not only to
better understand the links between local starburst objects and their
high-z counterparts, but also to gain valuable information on the
uncertainties associated with the derived properties of the stellar
populations in high-z star forming galaxies.

At this stage it it is clear that there are many unresolved issues
concerning the nature of ULIRGs. In particular:

\begin{itemize}
\item
Do the merger simulations adequately describe the observed properties
of ULIRGs?
\item
Do the properties of the stellar populations in ULIRGs correlate with
other properties of ULIRGs, such as infrared luminosity, interaction
class or spectral classification?
\item
What is the nature of the links between ULIRGs and other types of
object in the local universe, such as QSOs, radio galaxies and
elliptical galaxies?
\end{itemize}

Given their importance for studying the evolution of galaxies via
major gar-rich mergers, it is perhaps surprising that there have been
relatively few studies of stellar populations in ULIRGs. Many previous
studies have concentrated on studying the SEDs and mid-IR emission
line spectra of ULIRGs, with an emphasis on determining the dominant
heating source for the MFIR-emitting dust
\citep[e.g.][]{farrah03}. However, such studies are not capable of
determining the detailed properties of the stellar populations and how
they vary with spatial location in these complex systems.  Therefore,
order to address the key issues surrounding the evolution of ULIRGs we
have undertaken a programme of deep optical long-slit spectroscopic
observations of a substantial sample of nearby ULIRGs, aimed at
investigating their star formation histories. In Rodr\'iguez Zaur\'in
et al. (2009a), hereafter Paper I, we presented the sample, data
reduction and spectral syntehis modelling techniques. In this paper we
discuss the results presented in Paper I in the context of
evolutionary models for merging systems and the results obtained in
studies of stellar populations in high-z star forming galaxies.

Throughout this paper, we assume a cosmology with H$_{0}$ = 71 km
s$^{-1}$ Mpc$^{-1}$, $\Omega_{0} = 0.27$, $\Omega_{\Lambda} = 0.73$.

\section{Modelling the stellar populations: general results}

In Paper I we presented the results of spectral synthesis modelling of
the complete (CS) and the extended (ES) samples, including 26 and 36
ULIRGs respectively. The detailed properties of these samples are
described in that paper. To summarize, in the first place, we observed
a complete RA-limited and declination-limited sub-sample of the
\cite{Kim98a} 1 Jy sample of ULIRGs, with RAs in the range 12 $<$ RA
$<$ 1:30 h, declinations $\delta$ $>$ -23 degrees, and redshifts z~$<$
0.13. The redshift limit was chosen to ensure that the objects are
sufficiently bright for spectroscopic study, and also to keep the size
of the sample tractable for deep observations on 4m class
telescopes. We will refer to this sample of 26 objects as the complete
sample (CS, see Table 1 in Paper I for details of some of the the
properties of the ULIRGs in the CS). In addition, we also observed 10
objects outside the redshift range and/or the RA and Dec range of the
CS. With the aim of including a larger number of warm objects (only 8
warm ULIRGs are included in the CS), 7 of these 10 ULIRGs have warm
mid- to far-IR colours. The sample including the CS and these 10
additional objects will be referred to as the extended sample
(ES). Thus, the ES comprises 36 objects: 21 cool ULIRGs, and 15 warm
ULIRGs. Since the ES accounts better for the diversity within the
ULIRG phenomenon, and our statistical analysis shows no significant
differences between the properties of the stellar populations in the
ES and CS, we concentrate on the ES sample for the study presented in
this paper.

In order to perform a detailed study of the stellar populations in
ULIRGs, a total of 133 apertures was extracted for the objects in the
ES. The extraction apertures were selected from spatial cuts of the
2-D frames in the line-free continuum wavelength range 4400 --
4600~\AA, based on the visible extended structures and the requirement
that the apertures are large enough to have a sufficiently high S/N
ratio for further analysis. To compare the stellar
populations between the objects in our sample, apertures with a
metrical scale of 5kpc centred on the main nuclei were extracted for
all the objects in the extended sample (ES), including separate
extractions for multiple nuclei in individual sources. A second set of
apertures was then selected to sample the spatial features of those
objects in the ES showing tails, bridges and other diffuse
structures.

To perform the fits, we have used the CONFIT code (see
\citealt{Robinson00} and RZ08 for details). The CONFIT approach
consists of a direct fit of the overall continuum shape of the
extracted spectra using on a minimum $\chi^2$ technique
\citep{Tadhunter05,Rodriguez-Zaurin07}. CONFIT is based on a
``simplest model'' approach, i.e. we fit the minimum number of stellar
components required to adequately model the data.  Therefore, CONFIT
allows for a maximum of two stellar components plus a power-law in
some cases.  Throughout this paper, we define {\it young stellar
populations} (YSPs) as stellar components with ages t$_{YSP}$$\leq$ 2
Gyr, and {\it old stellar populations} (OSP) as components with ages
t$_{OSP}$$>$2 Gyr. With the aim of better describing each combination,
it is convenient to further sub-divide the YSPs into two groups: {\it
very young stellar populations} (VYSP):  stellar components with
ages t$_{\rm VYSP}$~$\leq$~0.1~Gyr; {\it intermediate-age young
stellar populations} (IYSP): stellar components with ages in the
range of 0.1 $<$ t$_{\rm IYSP}$ $\leq$ 2 Gyr.

Given that the properties of the stellar populations
were not know a priori, three modelling combinations were
investigated. The
results obtained for each of the combinations
are described in detailed in Paper
I, and are summarized as follows.
\begin{itemize}

\item {\bf Combination I.} As a first approach we used a two component
  model comprising a YSP (t$_{\rm YSP} \leq$ 2 Gyr) with varying
  reddening (0.0 $\leq E(B - V)\leq$ 2.0, increasing in steps of 0.1),
  along with an unreddened OSP of age 12.5 Gyr. Accounting for an
  old component covers the case in which one or more of the merging
  galaxies is early-type or bulge-dominated galaxy. Using this
  combination we obtained adequate fits ($\chi^{2}_{\rm red} \leq$
  1.0) to the overall shape of the SED for all but one (IRAS
  23060+0505, for which a power-law is required) of the extraction
  apertures modelled. However, Comb I fails to adequately fit the
  absoption detailed absorption -- particularly the CaII K feature --
  for $\sim$ 42\% (54) of the apertures modelled\footnote{As discussed
  in  paper I, some of this discrepancy may be due to the effect of
  ISM absorption on the CaII K resonance line.}.

\item {\bf Combination II.} This combination consists of three
  components: an OSP of age 12.5 Gyr and zero reddening, along with a
  YSP with variable reddening (0.0 $\leq E(B - V)\leq$ 2.0, increasing
  in steps of 0.1), and a power-law with a spectral index in the range
  $-15 < \alpha < 15$. The power-law is included to represent
  either scattered or direct AGN continuum component
  \citep{Tadhunter02}, or a highly reddened VYSP. Adequate fits
  ($\chi^{2} \leq$ 1.0) to the overall SED shapes were found for all
  of the apertures of all objects using this combination. On the other
  hand, this combination fails to fit the detailed absorption features
  for 10\% (13 apertures) of the apertures modelled. We find that the
  minimum precentage contribution of the OSP component is less that
  10\% (due to the uncertainties inherent in the modelling techniques
  percentage contributions $<$ 10\% are considered negligible) for
  most of the apertures. This result suggest that an OSP component is
  not required to fit the optical spectra of the majority of the
  objects in the ES.

\item {\bf Combination III.}  A drawback of Comb II is that, for those
cases in which the power-law is likely to represent a VYSP component,
it gives no information about the detailed properties (age and
reddening) of such a component. In addition, for ULIRGs as a class,
there is no reason, a priori, to expect a major contribution to the
optical emission from a 12.5 Gyr OSP. Such lack of an OSP is also
suggested by the results of Combination II. Therefore, in order to
explore the possibility of YSPs dominating the optical emission from
the objects we decided to use a combination consisting of two YSP
components: a IYSP with ages in the range 0.3 -- 2.0 Gyr, with $E(B -
V)$ values of 0.0, 0.2 or 0.4, plus a VYSP with age in the range 1 --
100 Myr and reddenings (0.0 $\leq E(B - V) \leq$ 2.0, increasing in
steps of 0.1). Adequate fits to both the continuum and the detailed
absorption features are obtained for all but 9 ($\sim$ 7\%) of the
extraction apertures modelled. In the minority of the cases for which
no adequate fits could be obtained using Comb III, either an OSP or a
power-law component was required in addition to a YSP.

\end{itemize} 

 Signifcantly, for the majority of apertures we find that  both
the SEDs and the detailed absorprion features can be adequately
modelled with all three combinations, although in most cases Comb
III provides the best overall fit. This demonstrates that it can be
difficult to derive a unique model solution even with high-quality,
wide spectral coverage optical spectra. Nonetheless, the fits
provide the following general results. 
\begin{itemize}

\item [-] {\bf Dominant young stellar populations.}  YSPs dominate the
optical light, while OSPs are relatively unimportant, in the
overwhelming majority of apertures modelled. In terms of the nuclear
apertures, the only exceptions are IRAS~13451+1232 5kpc, IRAS~1648NE
5kpc, IRAS~21208-0519 5kpcII, IRAS~23233+2817 5kpc and IRAS~23327+2913
5kpcII. Although there is some ambiguity in the cases of
IRAS~13451+1232 5kpc, IRAS~1648NE 5kpc, and IRAS~23233+2817 5kpc which
can be modelled either in terms of a dominant OSP plus a YSP (Comb I)
or a dominant ``old'' IYSP ($>$1~Gyr) plus a VYSP (Comb III), we can
only adequately model IRAS~21208-0519 5kpcII and IRAS~23327+2913
5kpcII with Comb I models that include an OSP. Note that, in the case
of IRAS~23327+2913 5kpcII, it is possible to model the optical
spectrum using an OSP, without any requirement for a YSP. Apart from
these exceptions, for most objects the results are consistent with the
idea that the precursor galaxies are gas-rich spiral galaxies.
\item [-] {\bf Complex star formation histories.}  Consistent with our
previous results for the nearby ULIRG Arp220
\citep{Rodriguez-Zaurin08}, we find that most apertures of most
objects are best fitted with models that include a combination of VYSP
and IYSP (i.e. Comb III). This suggests that the recent star formation
histories of most ULIRGs are complex, with at least two star formation
epochs, even if the most recent (represented by the VYSPs) often
dominates the optical light.
\end{itemize}

In the following sections we consider the properties of the YSP in
ULIRGs in more quantitative detail, and investigate whether they are
correlated with other properties of these merging systems.

\subsection{Characteristic properties of the YSPs 
in the nuclear and extended regions.}

As described in Paper I, Comb I (YSP + OSP) provides useful estimates
of the ``luminosity weighted'' ages, reddenings and percentage
contributions of the dominant YSPs within each galaxy. Providing a
single estimate (with an associated uncertainty) for the parameters
associated with the YSPs in each aperture, this combination is
particularly useful for performing a statistical analysis of the
properties of the stellar populations in the CS and the ES samples.

\begin{figure*}
\begin{minipage}{\textwidth}
\begin{tabular}{cc}
\hspace*{1cm}\psfig{file=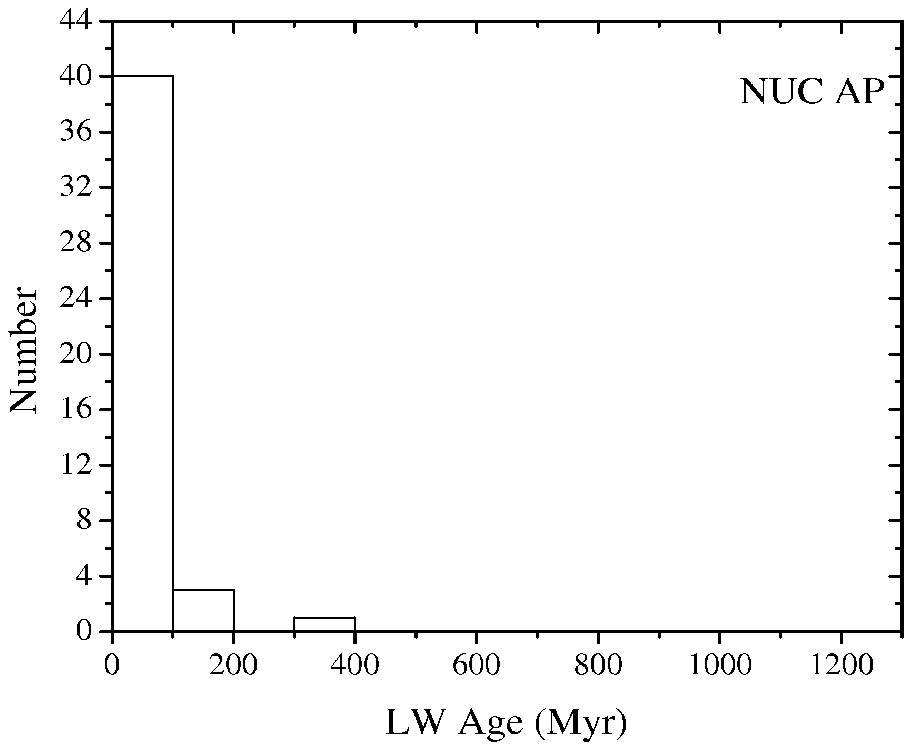,width=6.8cm,angle=0.}&
\psfig{file=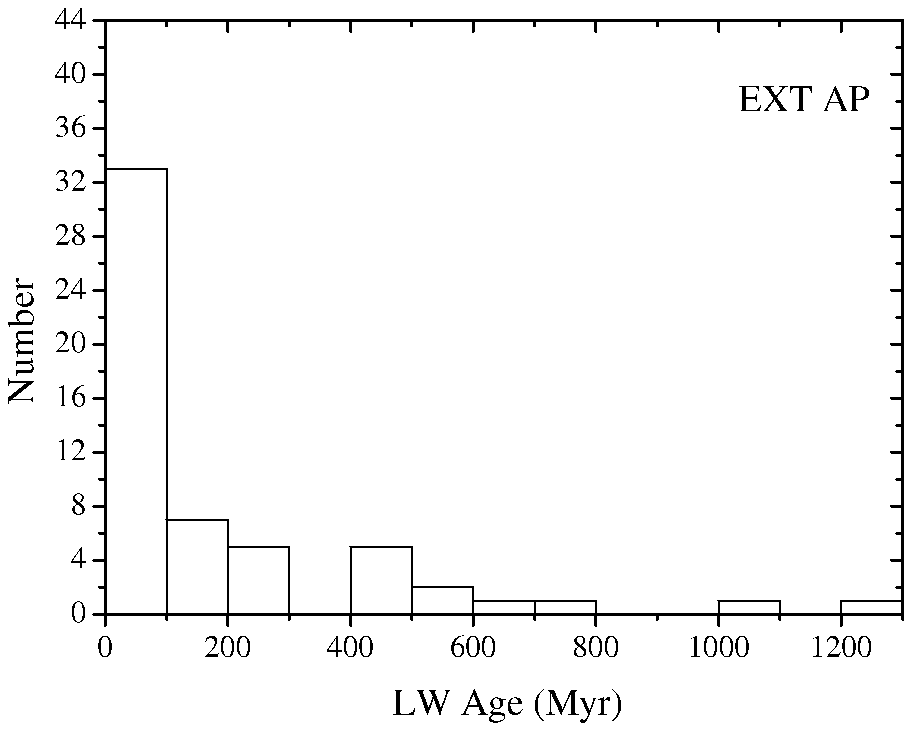,width=7.0cm,angle=0.}\\
\hspace*{1cm}\psfig{file=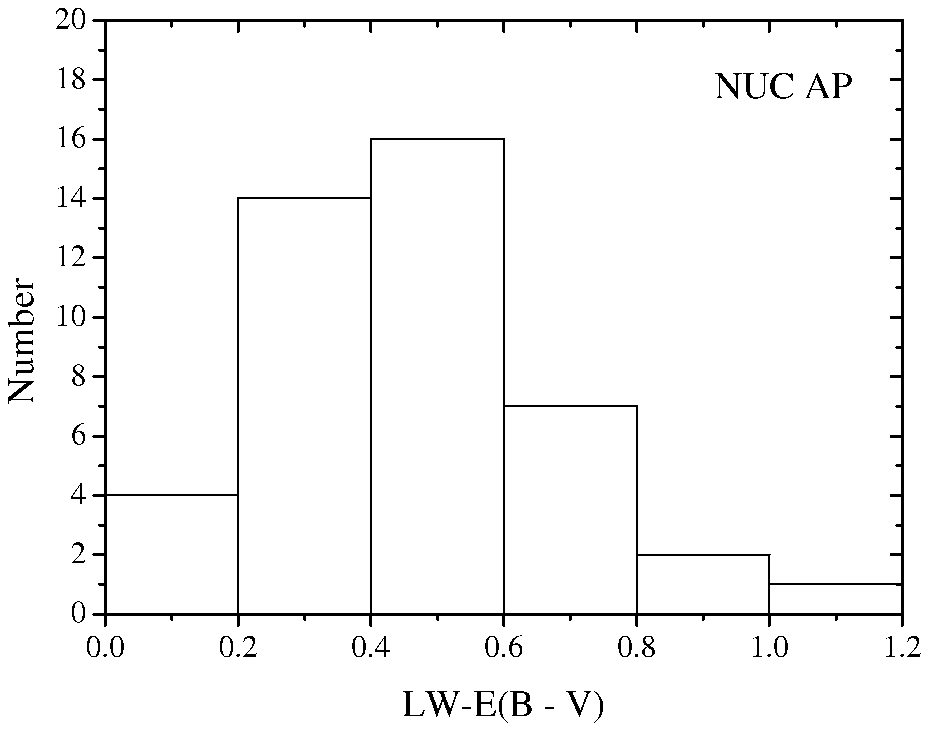,width=7.0cm,angle=0.}&
\psfig{file=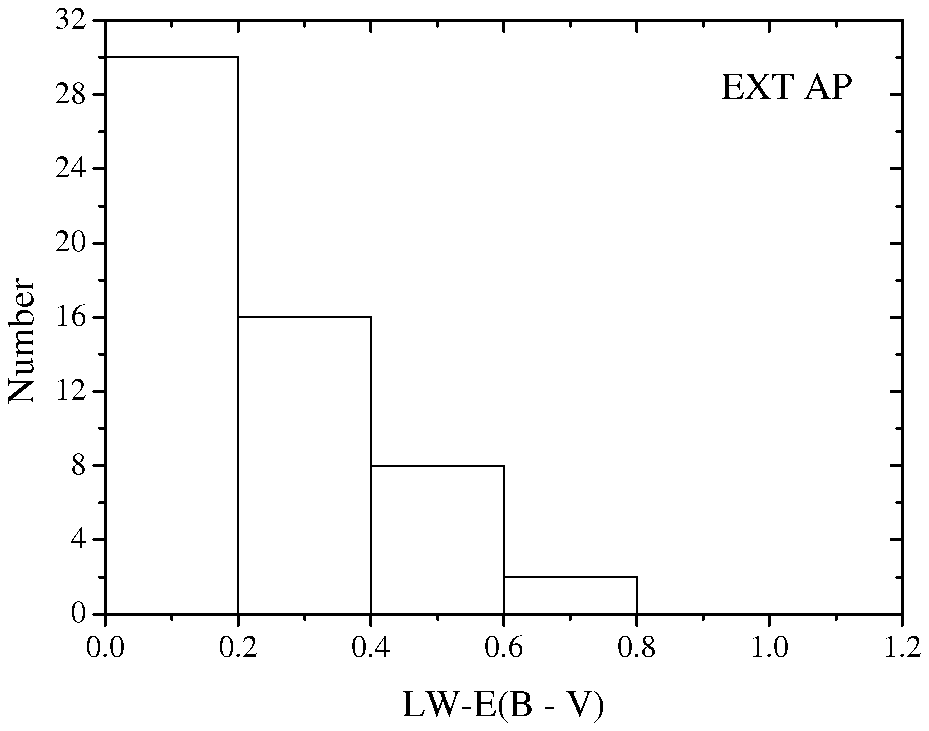,width=7.0cm,angle=0.}\\
\hspace*{1cm}\psfig{file=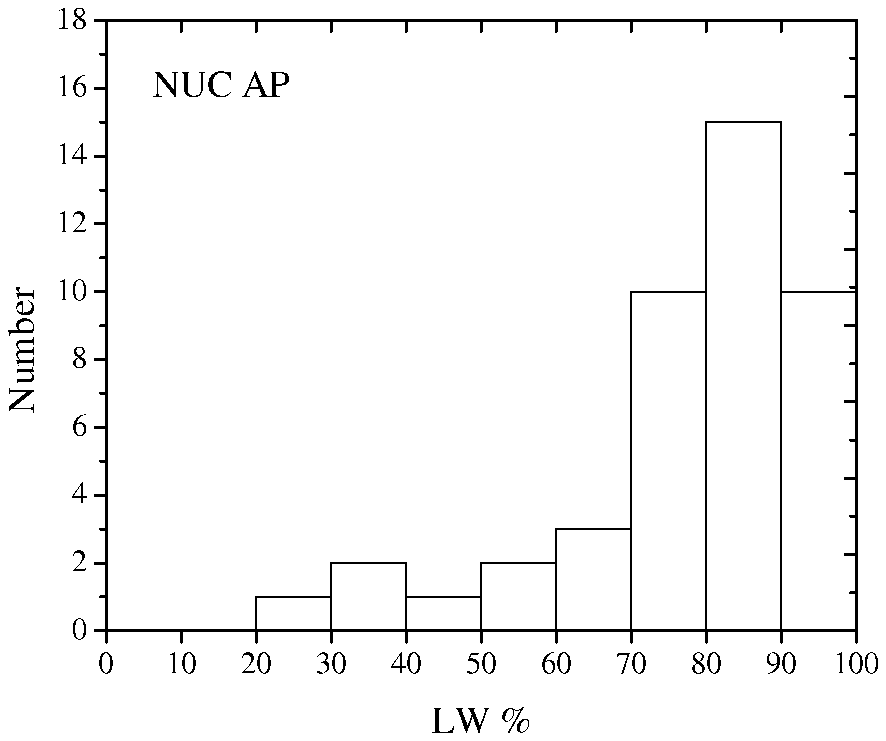,width=7.0cm,angle=0.}&
\psfig{file=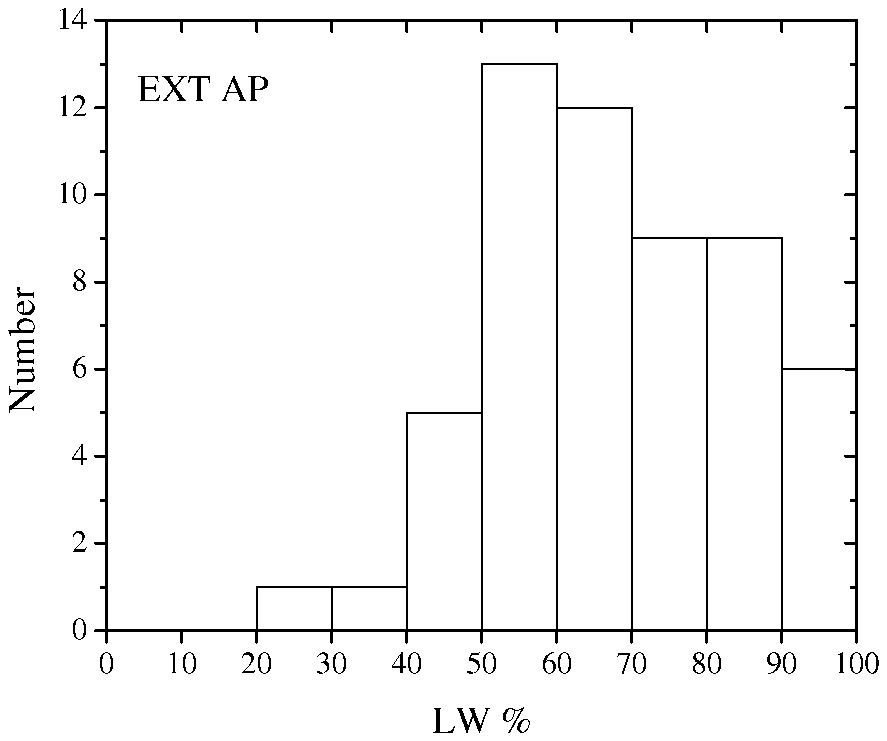,width=7.0cm,angle=0.}\\
\end{tabular}
\caption{Histograms showing the LW-age (top), LW-E(B - V) (middle),
and LW-\% (bottom) contribution distributions for the nuclear (left) 5
kpc and the extended (right) apertures. The results of Comb I show
that YSPs are present at all locations in the galaxies for the
majority of the objects in the ES, with a clear cut-off in YSP ages at
100 Myr in the case of the nuclear regions. The figure also suggests
that the YSPs in the nuclear regions of the galaxies, coinciding with
higher reddening, represent a more important contribution to the flux
at optical wavelengths than those in the extended regions.}
\label{fig:combI_hist_ES}
\end{minipage}
\end{figure*}

Figure \ref{fig:combI_hist_ES} shows the distributions of the average
luminosity-weighted ages (LW-age), reddenings (LW-$E(B - V)$) and
percentage contributions to the flux in the normalising bin (LW-\%) of
the YSPs in the nuclear and extended apertures. This figure has been
made using  34 of the 36 objects in the ES. For two of the three
objects classified as Sy1 (IRAS 15462-0450 and IRAS 21219-1757) no
extended apertures were extracted and, therefore, they are not
included in the analysis presented here. In the case that one of the
extended apertures samples a similar region to the 5 kpc aperture,
only the latter is used for the plots (for example, ApE and 5 kpc in
the case of Mrk 273). In the case of Arp 220, only two of the 24
apertures were included, in order to avoid giving too much weight to
the extended apertures in a single object. These are the 5 kpc
aperture and AP$_{\rm TOTAL}$ for PA 75*.  Overall, a total of 100
of the 128 apertures modelled for the ES were eventually used for the
figures.

\begin{table}
\centering
{\small
\begin{tabular}{@{}llll@{}}
\hline
&&Mean & Median\\
\hline
       &LW-age (Myr)& 66 & 60\\
NUC AP &LW-E(B - V)& 0.4  & 0.4\\
       &LW-\%      & 77  & 82\\
\hline
       &LW-age (Myr)& 203  & 80\\
EXT AP &LW-E(B - V)& 0.2 & 0.2\\
       &LW-\%      & 63& 66\\
\end{tabular}}
\caption{Mean and median values for the parameters associated with the
YSPs for both the nuclear and extended regions of the objects in the
ES, obtained using Comb I.}
\label{tab:Stat_CombI_ES} 
\end{table}
\begin{table}
\centering
{\small
\begin{tabular}{@{}llll@{}}
\hline
&D &Confidence\\
&&Level\\
\hline
LW-age      & 0.325& 99\%\\ 
LW-E(B - V) & 0.50 & $>$ 99.9\%\\
LW-\%       & 0.383& $>$ 99.9\%\\   
\end{tabular}}
\caption{Results of the K-S test for the significance of differences
between the distributions in the nuclear and the extended regions, of
the LW-age, LW-E(B - V) and LW-\% for the ES and using Comb I. Col
(2): the maximum deviation between the two cumulative distribution
functions. Col (3): the probability that the difference between the
distribution of the nuclear and the extended regions is significant.}
\label{tab:K-S_CombI_ES} 
\end{table}

The upper panel of Figure \ref{fig:combI_hist_ES} shows the presence
of a clear upper limit of 100 Myr for the LW-ages of the YSPs in the
nuclear apertures; only $\sim$ 9\% of the nuclear apertures (4 of
the 44 nuclear apertures used) have average LW-ages older than this
value. In contrast, $\sim$41\% of the extended apertures (23 of
the 56 extended apertures used) have average LW-age values older than
100 Myr, and as old as 1.2 Gyr. In terms of reddening, Figure
\ref{fig:combI_hist_ES} (middle) suggests the presence of higher
reddening (LW-$E(B - V)$) values in the nuclear regions of the
galaxies compared with the extended regions. We also find that the
YSPs represent the dominant contribution (LW-\% $\geq$ 50\%) to the
optical emission for all but 11 (11\%) of the apertures
used (4 nuclear apertures and 7 extended apertures, Figure
\ref{fig:combI_hist_ES}; bottom panel). However, the YSP make a more
important contribution to the optical light in the nuclear
regions. 

Table \ref{tab:Stat_CombI_ES} shows the mean and median
values of the different parameters associated with the YSPs. To
investigate whether the distributions of the YSPs properties in the
nuclear and extended regions are indeed different, we used the
Kolmogorov-Smirnov (K-S) two sample test, and the results are
presented in Table \ref{tab:K-S_CombI_ES}. We find that the
differences between the LW-age, LW-$E(B - V)$ and LW-\% distributions
for the YSPs in the nuclear and extended regions of the galaxies are
significant with a confidence level of $\gsim$99\% (i.e. the
probability that the distributions are drawn at random from the same
parent distribution is $<$1\%).

\begin{figure*}
\begin{minipage}{\textwidth}
\begin{tabular}{cc}
\hspace*{1cm}\psfig{file=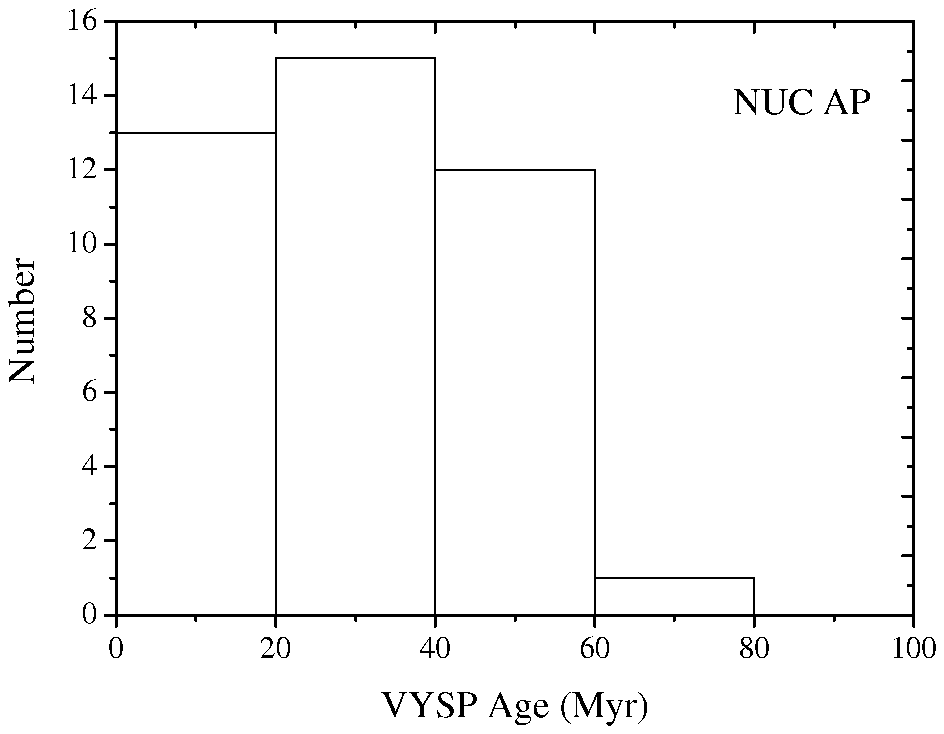,width=7.0cm,angle=0.}&
\psfig{file=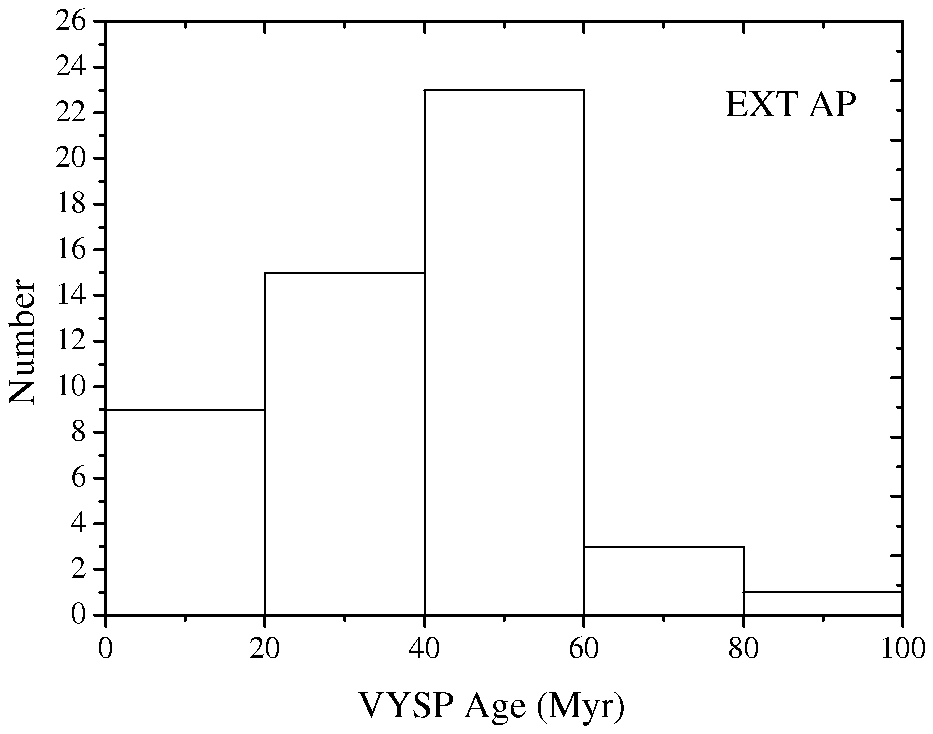,width=7.0cm,angle=0.}\\
\hspace*{1cm}\psfig{file=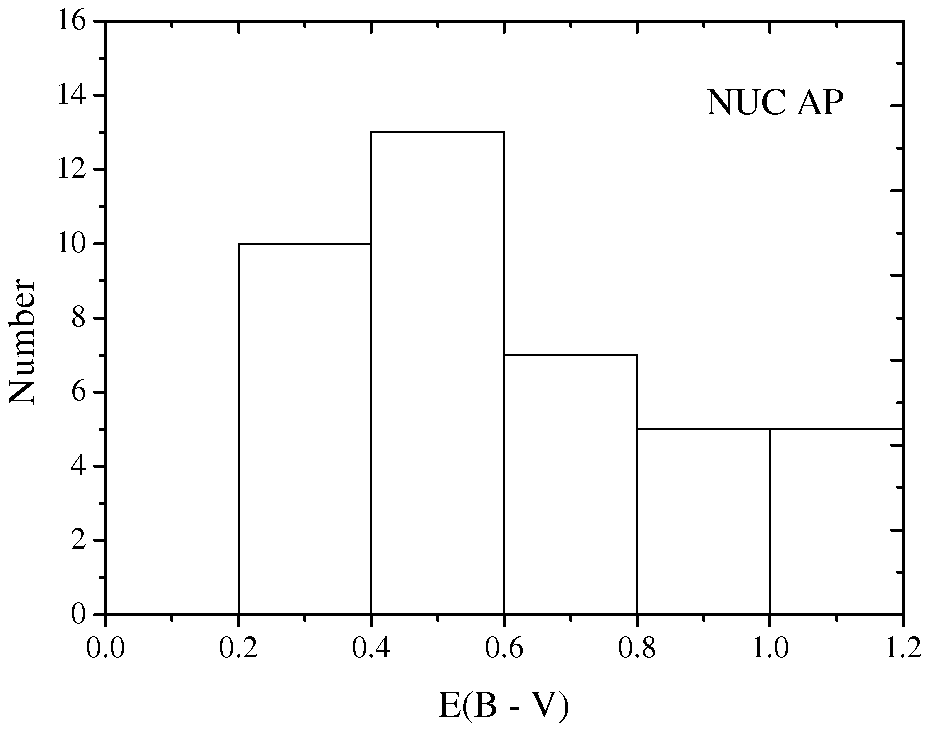,width=7.0cm,angle=0.}&
\psfig{file=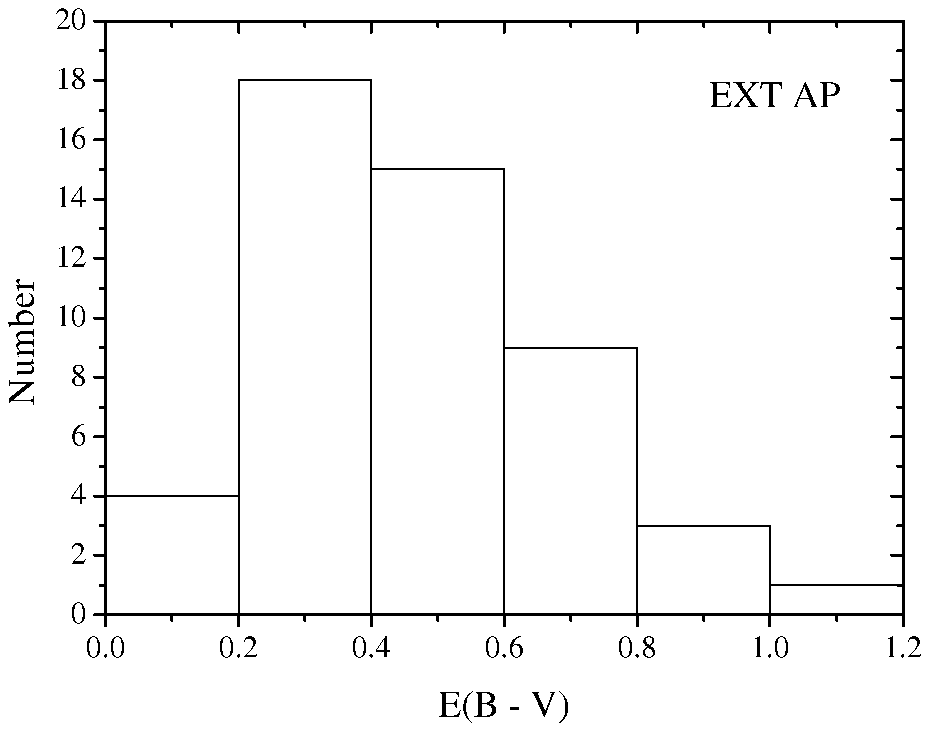,width=7.0cm,angle=0.}\\
\hspace*{1cm}\psfig{file=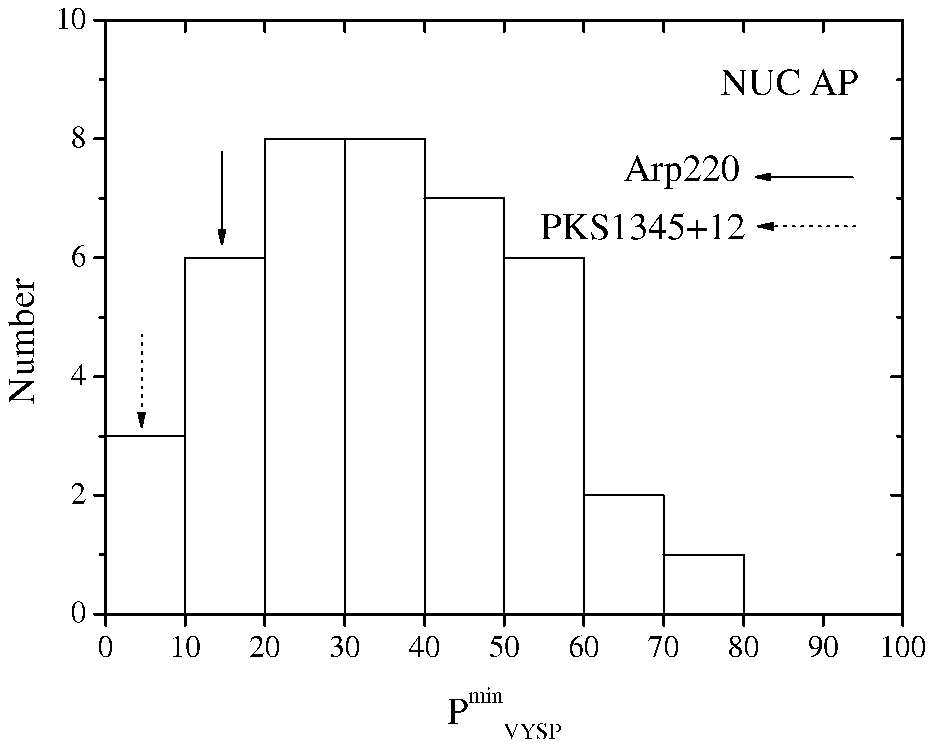,width=7.0cm,angle=0.}&
\psfig{file=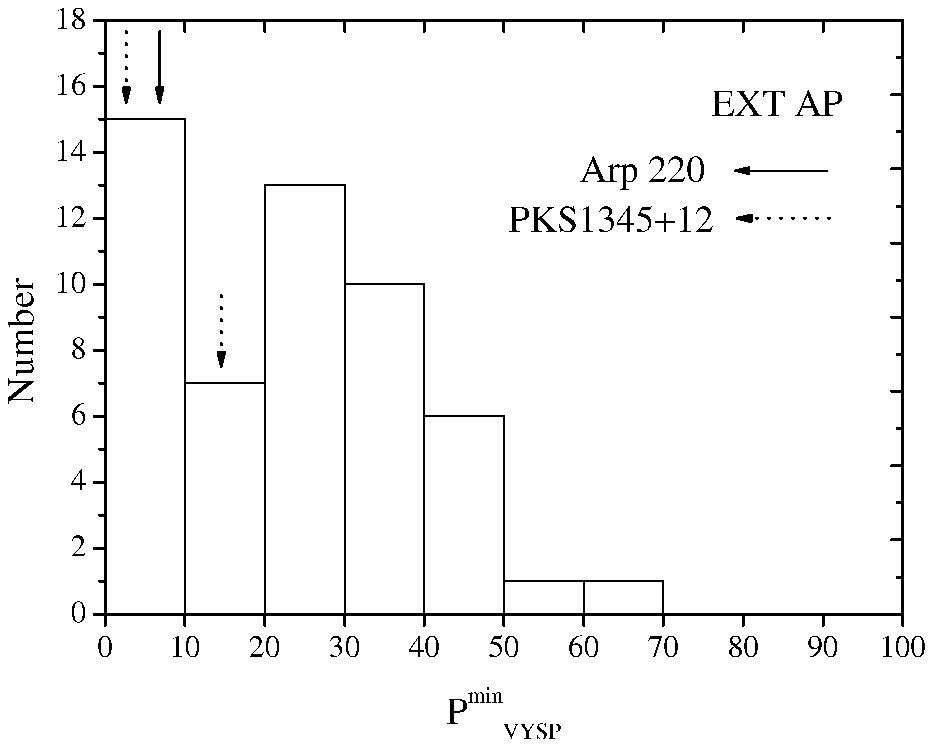,width=7.0cm,angle=0.}\\
\end{tabular}
\caption{Histograms showing the distributions of the average VYSP-Age
(top), E(B - V) values (middle) and the minimum percentage
contribution to the flux in the normalising bin (P$^{\rm min}_{\rm
VYSP}$, bottom) for the nuclear 5 kpc (left) and the extended (right)
apertures, using the ES and Comb III. The locations of Arp 220 and
PKS1345+12 are also shown in the figure.}
\label{fig:combIII_hist_ES}
\end{minipage}
\end{figure*}
\begin{table}
\centering
{\small
\begin{tabular}{@{}llll@{}}
\hline
&&Mean & Median\\
\hline
       &VYSP-Age (Myr)     & 31 & 30\\
NUC AP &VYSP-E(B - V)      & 0.6 & 0.6\\
       &P$^{\rm min}_{\rm VYSP}$ & 37 & 39\\
\hline
       &VYSP-Age (Myr)     &39 & 50\\
EXT AP &VYSP-E(B - V)      &0.4 & 0.4\\
       &P$^{\rm min}_{\rm VYSP}$ &24 & 25\\
\end{tabular}}
\caption{Mean and median values for the parameters associated with the
VYSPs, for both the nuclear and extended regions of the objects in the
ES, obtained using Comb III.}
\label{tab:Stat_CombIII_ES} 
\end{table}
\begin{table}
\centering
{\small
\begin{tabular}{@{}llll@{}}
\hline
&D &Confidence\\
&&Level\\
\hline
VYSP-Age (Myr)     & 0.27 & 95\%\\ 
VYSP-E(B - V)      & 0.27 & 95\%\\
P$^{\rm min}_{\rm VYSP}$ & 0.35 & 99.5\%\\   
\end{tabular}}
\caption{Results of the K-S test for the significance of differences
between the distributions in the nuclear and the extended regions, of
the VYSP-Age, VYSP-E(B - V) and P$^{\rm min}_{\rm VYSP}$ for the ES,
using Comb III. Col (1): the maximum deviation between the cumulative
distribution functions Col (2): the probability that the difference
between the distribution of the nuclear and the extended regions is
significant.}
\label{tab:K-S_CombIII_ES} 
\end{table}

To further investigate the properties of the stellar populations, we
performed a similar study to that presented above using the results
of Comb III (VYSP + IYSP). Figure \ref{fig:combIII_hist_ES} shows the
distributions of the average values for the VYSPs age, reddening and
the minimum percentage contribution to the flux in the normalising bin
(P$^{\rm min}_{\rm VYSP}$), for both the nuclear and the extended
regions. From a visual inspection of the histograms, the modelling
results suggest the presence of ``younger'' and more reddened VYSPs in
the nuclear regions of the objects. Moreover, the lower panel in
Figure \ref{fig:combIII_hist_ES} is designed to show the importance of
the VYSPs within the different regions of the galaxies sampled by the
apertures. The figure suggests that VYSPs are more important in the
nuclear regions of the galaxies. Table \ref{tab:Stat_CombIII_ES}
presents the mean and median values for the VYSP-Age, VYSP-E(B - V)
and P$^{\rm min}_{\rm VYSP}$, while Table \ref{tab:K-S_CombIII_ES}
shows the results of the K-S test obtained using Comb III. {We find
that the difference between the VYSPs age and E(B - V) distributions
in the nuclear and the extended regions is significant with a
confidence level of 95\%. In the case of P$^{\rm min}_{\rm VYSP}$, the
confidence level is 99.5\%. Overall, the results are consistent with
those of Comb I, but the level of significance is lower. 

The lower panel of Figures \ref{fig:combIII_hist_ES} also shows the
locations of IRAS 13451+1232 (PKS1345+12) and Arp 220, the two objects
for which detailed studies were presented in \cite{Rodriguez-Zaurin07}
and \cite{Rodriguez-Zaurin08}. Arp 220 is frequently used as the
archetype of ULIRGs as a class. However, the modelling results
indicate that the YSPs in this object are not typical of the objects
ULIRGs in general. While significant VYSPs are detected at all locations
of the galaxies sampled by the apertures, in the case of
Arp 220 the contribution of such populations is relatively small, apart
from in the nuclear region of the galaxy. Furthermore, in this
object, a uniform IYSP of age 0.5 -- 0.9 Gyr is found at all
locations sampled by the apertures.

In the case of PKS 1345+12, some peculiarities might be expected, since
this is the only object in our sample classified as radio galaxy. For
this object, either OSPs or dominant ``old'' IYSPs (1 - 2 Gyr) are
required to model the data (see Paper I for details); the contribution
from VYSPs is relatively small. Independent of the combination of
stellar populations assumed, this galaxy is the most massive amongst the
objects in our sample (see Table 4 in Paper I).

Overall, our
modelling results indicate that the VYSPs in the nuclear regions of
the galaxies make a more significant contribution to the optical light
than those of the extended regions. In terms of age and reddening, the
VYSPs located in the nuclear regions tend to be younger and more
reddened, although further analysis is required to confirm these
results. This results will be discussed in the context of the merger
simulations in Section 6.

\subsection{Correlations between the properties of the YSP and other
ULIRG properties}

Given that certain of the general properties of ULIRGs (e.g. IR luminosities,
mid- to far-IR colours, optical morphology and spectral class) may 
change as the systems evolve along the merger sequence, it is 
interesting to determine whether such properties correlate with 
those of the YSP determined from spectral synthesis modelling.

With the aim of investigating whether the more luminous objects have
younger, redder or more important VYSPs, Figure \ref{fig:combIII_SC}
shows the average VYSP age, reddening (E(B - V)) and minimum
percentage contribution (P$^{\rm min}_{\rm VYSP}$) obtained for the
nuclear 5 kpc apertures, plotted against the log of the IR-luminosity
(from Kim \& Sanders 1998). Note that, since it was not possible to
model the Sy1-ULIRGs (IRAS 12540+5708, IRAS 15462-0450 and IRAS
21219-1757, see Paper I for details), no Sy1 galaxies are shown in the
figure. We find that the properties of
the VYSPs are independent of the IR-luminosities of the sources. 

In addition to looking for trends with luminosity we can also examine
whether the stellar properties depend on the optical spectral
classification. As can be seen from Figure \ref{fig:combIII_SC}, where
the points are labelled according to spectral classification, we see
no clear trend with optical type. Table \ref{tab:Statistics_SC} shows
the mean and median values of the VYSP age, reddening and P$^{\rm
min}_{\rm VYSP}$ for the different types of object i.e. HII-, LINER-
and Sy2-ULIRGs. If, for example, there exists an evolutionary link
between the HII- or LINER-ULIRGs and the Sy2-ULIRGs (Sanders et al,
1988b), the latter representing a more advanced stage in an
evolutionary sequence, one might expect the Sy2-ULIRGs to have older
ages.  No trends are observed in Figure \ref{fig:combIII_SC} for the
different types of ULIRGs. A first interpretation is that there is no
evolutionary link between HII-, LINER- and Sy2-ULIRGs. However, it is
also possible that the transition between one type and the other
occurs over a shorter timescale than the typical timescale of the
enhancement of the star formation activity in the final stages of the
merger event as the nuclei coalesce \cite[$\sim$ 100
Myr,][]{Barnes96}. In that case, no trend is expected to be observed.

\begin{figure}
\centering
\hspace{0.0cm}
\psfig{figure=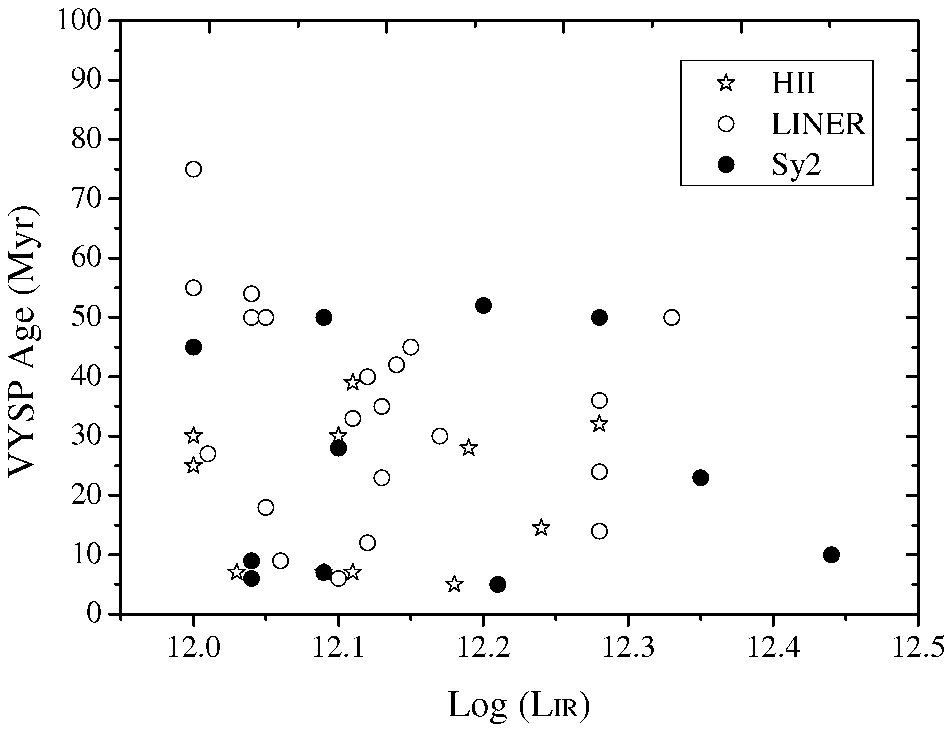,width=8.5cm,angle=0.}\\
\psfig{figure=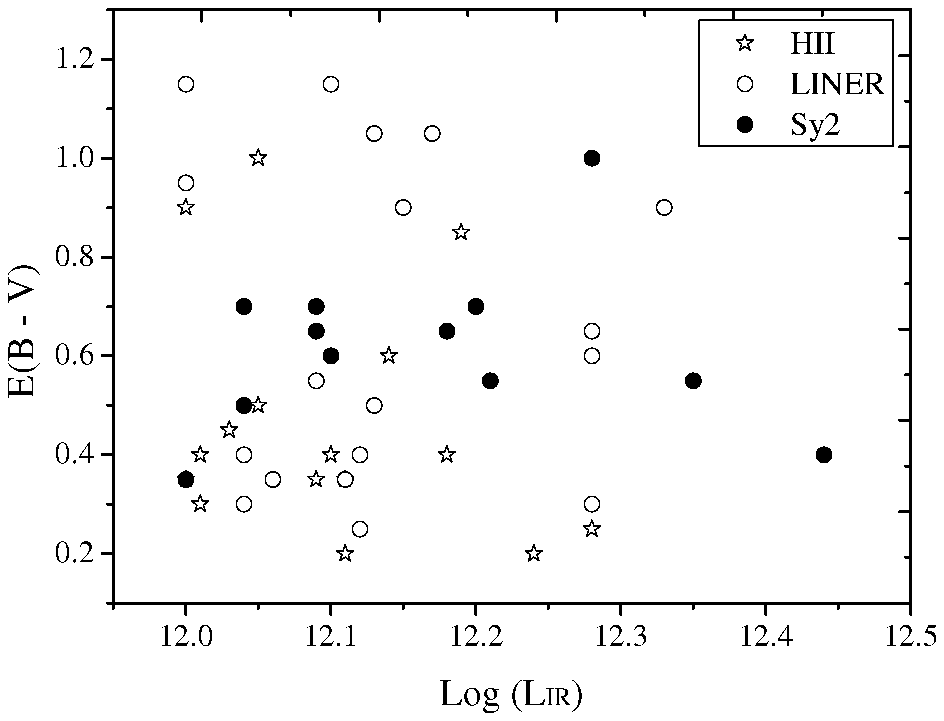,width=8.5cm,angle=0.}\\
\psfig{figure=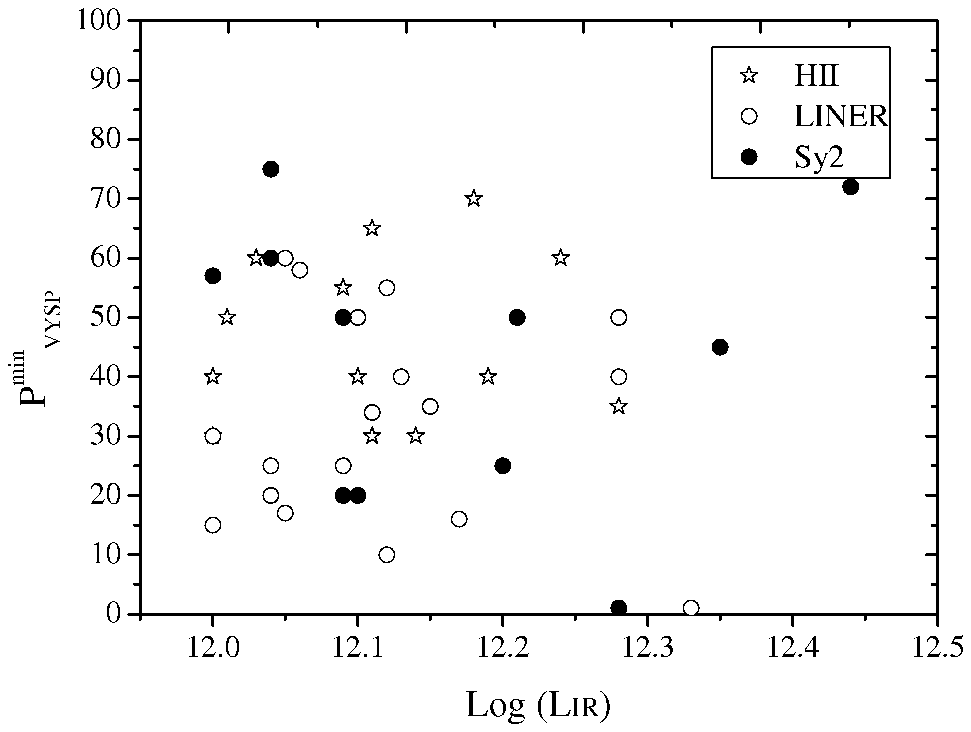,width=8.5cm,angle=0.}
\caption[Average VYSP-age, E(B - V), and P$^{\rm min}_{\rm VYSP}$
plotted against the infrared luminosities. Points labelled based on
the spectral classification]{Average VYSP age, E(B - V), and P$^{\rm
min}_{\rm VYSP}$ plotted against Log(L$_{\rm IR}$). The nuclear 5 kpc
apertures have been used for these plots. The points are labelled
based on the Veilleux et al. (1999) spectral classifications. No
significant correlations are observed.}
\label{fig:combIII_SC}
\end{figure}

\begin{table}
\centering
{\small
\begin{tabular}{@{}lllll@{}}
\hline
&&Mean & Median&\\
\hline
     & VYSP-Age (Myr)      &20  &25\\
HII & VYSP-E(B -V)        &0.5&0.4\\
     & P$^{\rm min}_{\rm VYSP}$  &46&40\\
\hline
     & VYSP-Age (Myr)      &35  &35\\
LINER& VYSP-E(B -V)        &0.6&0.6\\
     & P$^{\rm min}_{\rm VYSP}$  &34  &34\\
\hline
     & VYSP-Age (Myr)      &28  &25\\
Sy2  & VYSP-E(B -V)        &0.6&0.6\\
     & P$^{\rm min}_{\rm VYSP}$  &43  &50\\
\hline
\end{tabular}}
\caption[Mean and median values of the VYSP-Age, VYSP-E(B -V) and 
P$^{\rm min}_{\rm VYSP}$ for the HII, LINER and Sy2-ULIRGs]{Mean and
median values of the VYSP-Age, VYSP-E(B -V) and P$^{\rm min}_{\rm
VYSP}$ for the HII, LINER and Sy2-ULIRGs, for the nuclear 5 kpc
apertures.}
\label{tab:Statistics_SC} 
\end{table}

We have also investigated the presence of correlations between
the properties of the VYSPs and other properties of ULIRGs such as the
Veilleux et al. (2002) interaction class, or the MFIR colour ratio
{\it f\/}$_{25}$/{\it f}$_{60}$\footnote{The quantities {\it
f\/}$_{25}$ and {\it f}$_{60}$ represent the {\it IRAS} flux densities
in Jy at 25$\mu$m and 60$\mu$m.}. The conclusions reached are
identical, i.e. no evidence for correlations or trends is found.

As well as the explanation already discussed above, possible reasons
for the lack of correlations between the YSP properties and other
properties of ULIRGs include the following:

\begin{itemize}

\item{\bf Modelling technique}: the outputs of the modelling technique
  used for the analysis presented in this section, as described in detail
  in Paper I, are ranges of age, reddening and percentage contribution
  of both IYSPs and VYSPs. Therefore, part of the scatter in the YSP
  properties could be due to the uncertainties inherent in the
  modelling technique; such uncertainties could hide underlying trends
  and patterns in behaviour.

\item{\bf Other variables}: there are other variables such as, for example, 
the gas contents of the parent galaxies or the geometry of the merger,
that could have a more important impact on the properties of the VYSPs
in ULIRGs than the spectral classification, MFIR colours, MFIR
luminosities, or the interaction class.
\end{itemize}

Interestingly, this apparent lack of correlations between the various
properties of ULIRGs has been found by other authors. Veilleux et
al. (1999) found no correlation between the emission line reddenings
and other properties of ULIRGs, such as the MFIR colour ratio {\it
f\/}$_{25}$/{\it f}$_{60}$, for the 108 objects in their
sample. Moreover, \cite{Farrah01} carried out an {\it HST}-WFPC2
V-band imaging study of a sample of 23 ULIRGs and found no correlation
between the IR-luminosity and the morphologies of the objects in their
sample.

\section{The masses of the YSPs}

Table 4 in Paper I presents estimates for the total stellar masses
associated with the different stellar components for the objects in
the ES. Figure \ref{fig:Hist_Mass} shows the distribution of the total
stellar masses presented in that table (accounting for all VYSP, IYSP
and OSP if present). Note that, in the cases of IRAS 14394+5332, IRAS
17028+5817 and IRAS 23327+2913, the total stellar mass was estimated
for the two nuclei individually and, for the purpose of the figure,
the masses of the two nuclei were added. The figure shows that only 6
objects in our sample --- IRAS 00188-0856, IRAS 13451+132, IRAS
14348-1447, IRAS 17179+5444, IRAS 23233+2819 and IRAS 23327+2913 ---
have stellar masses M$_{\rm stellar}$ $>$ 1 $\times$ 10$^{11}$
M$_{\odot}$. Although two of these objects --- IRAS 1345+132, IRAS
17179+5444 --- are spectroscopically classified as Sy2 galaxies in the
optical, overall there is no clear-cut distinction between Seyfert and
non-Seyfert ULIRGs in terms of their stellar mass; many Seyfert ULIRGs
have stellar masses that are relatively modest -- at or below the
median for the sample as a whole.  Thus our results provide no clear
support for the idea that AGN activity in ULIRGs is more likely to be
associated with relatively early-type galaxies with massive bulges, as
appears to be the case in large samples of galaxies drawn from the
Sloan Digital Sky Survey (SDSS) \citep{kauffmann04,heckman04}.

\begin{figure}
\centering
\hspace{0.0cm}
\psfig{figure=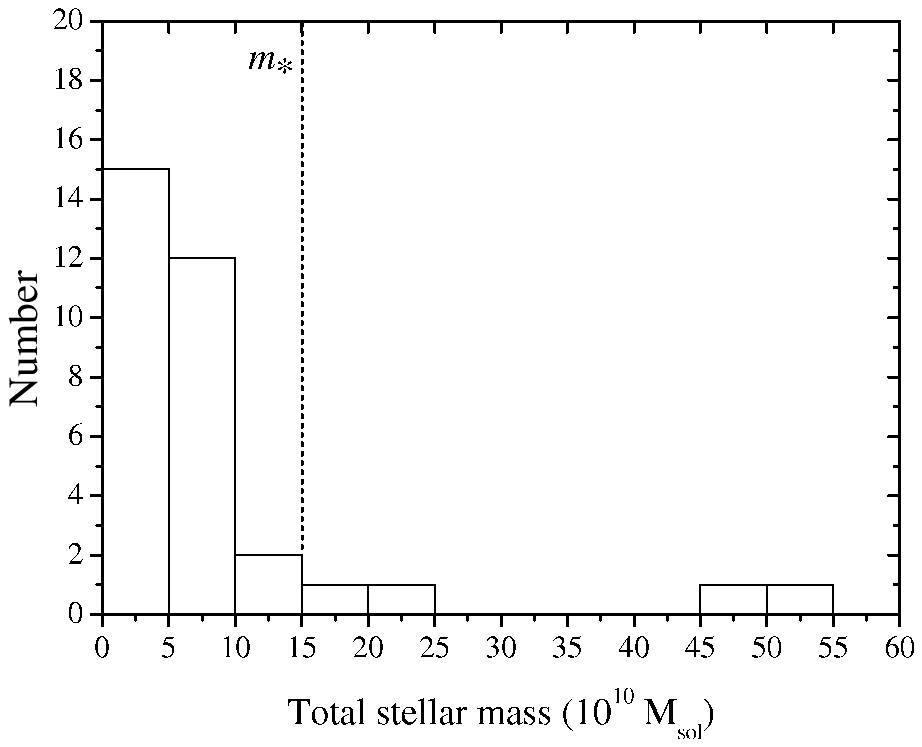,width=8.cm,angle=0.}
\caption[Histogram showing the distribution of the total stellar 
mass values obtained for the ES]{Histogram showing the distribution of
the total stellar mass values presented in Table \ref{tab:Mass_Lbol_mstar}
for the objects in the ES. The figure shows that only 6 of the objects
have stellar masses M$_{\rm stellar} >$ 1 $\times 10^{11}$
M$_{\odot}$.}
\label{fig:Hist_Mass}
\end{figure}

\begin{table*}
\centering
{\small
\begin{tabular}{@{}lcccc@{}}
\hline
Object Name& M$_{\rm stellar}$ & M$_{\rm dyn}^{1}$ & M$_{\rm dyn}^{2}$ & M$_{\rm dyn}^{3}$ \\    
IRAS       &  ($m_{\ast}$) & ($m_{\ast}$) & ($m_{\ast}$) & ($m_{\ast}$) \\
\hline 
00091--0738 &   0.33  & ...  & ...  & ... \\
00188--0856 & 	0.72  & ...  & ...  & ... \\
01004--2237 & 	0.11  & ...  & ...  & 0.07\\
08572+3915  & 	0.20  & ...  & 0.13 & ... \\
10190+1322  & 	0.28  & ...  & ...  & 0.93\\
10494+4424  & 	0.37  & ...  & ...  & ... \\
12072--0444 & 	0.26  & ...  & ...  &...  \\
12112+0305  & 	0.52  & ...  & 0.70 & 0.40\\
12540+5708  &	 ...  & ...  & ...  & ... \\
13305--1739 & 	0.28  & ...  & ...  & ... \\
13428+5608  & 	0.53  & 2.0  & 0.6 - 1.1 & 0.88\\
13451+1232  & 	3.60  & ...  & ...  & ... \\
13539+2920  & 	0.24  & ...  & ...  & ... \\
14060+2919  & 	0.40  & ...  & ...  &...  \\
14252--1550 & 	0.38  & ...  & ...  & ... \\
14348--1447 & 	0.77  & 1.0  & 1.1  & 1.50\\
14394+5332  & 	0.44  & ...  & ...  & ... \\
15130--1958 & 	0.28  & ...  & ...  & 0.51\\
15206+3342  & 	0.60  & ...  &  0.3 &...  \\
15327+2340  & 	0.25  & 0.3  &  0.3 & 0.26\\
16156+0146  & 	0.22  & ...  & ...  & ... \\
16474+3430  & 	0.61  & ...  & ...  & ... \\
16487+5447  & 	0.67  & ...  & ...  & ... \\
17028+5817 & 	0.40  & ...  & ...  & ... \\
17044+6720  & 	0.19  & ...  & ...  & ... \\
17179+5444  & 	1.24  & ...  & ...  & ... \\
20414-1651  & 	0.25  & ...  & ...  & 0.52\\
21208--0519 &   0.68  & ...  & ...  & ... \\
22491--1808 &   0.20  & ...  & ...  & 0.68\\
23060+0505  &   0.49  & ...  & ...  & ... \\
23233+2817  &   1.64  & ...  & ...  & ... \\
23234+0946  &   0.34  & ...  & ...  & ... \\
23327+2913  &   3.37  & ...  & ...  & ... \\
23389+0303  &   0.15  & ...  &  ... & ... \\
\end{tabular}}
\caption[Estimated stellar masses in units of $m_{\ast}$]{{\small
Estimated total masses associated with the stellar populations (YSPs,
including VYSPs and IYSPs, and OSPs if present) detected at optical
wavelengths, expressed in units of $m_{\ast}$ ($m_{\ast}$ = 1.4
$\times 10^{11}$ M$_{\odot}$). Due to the powerful AGN emission, it
was not possible to estimate the total mass for the three objects in
our sample classified as Sy1 galaxies (IRAS 12540-5708, IRAS
15462-0405 and IRAS 21219-1757).\newline $^{1}$ Dynamical mass from
\cite{Tacconi02}. \newline $^{2}$ Dynamical mass from
\cite{Colina05}. \newline $^{3}$ Dynamical mass from
\cite{Dasyra06a,Dasyra06b}}.}
\label{tab:Mass_Lbol_mstar} 
\end{table*}

In order to compare the results found here with other studies, Table
\ref{tab:Mass_Lbol_mstar} shows the total mass values in units of
$m_{\ast}$, defined as the break in the measured mass function
obtained from a large sample of 17,173 galaxies drawn from a
combination of the Two Micron All Sky survey (2MASS) Extended Source
catalogue and the 2dF galaxy redshift survey \cite[$m_{\ast}$ = 1.4
$\times 10^{11}$ M$_{\odot}$;][adapted to our cosmology]{Cole01}. For
the ES sample as a whole, the stellar masses are in the range
0.1~$m_{\ast}$ $\leq$ M$_{\rm stellar}$ $\leq$ 3.6 $m_{\ast}$, with a
mean value of 0.6 $m_{\ast}$ and a median of 0.4
$m_{\ast}$. Aproximately 80\% of the objects in the ES (27) have total
stellar masses M$_{\rm stellar} < $ 1 $\times 10^{11}$ M$_{\odot}$
(0.71 $m_{\ast}$). Also shown in the table are the dynamical mass
estimates (including stars, gas and dark matter) from
\cite{Tacconi02}, \cite{Colina05} and \cite{Dasyra06a,Dasyra06b} for
the objects in common with the ES.  The values presented in the table
are for the entire systems, including the extended regions of the
galaxies and adding the dynamical masses of the two nuclei for the
double nucleus systems. To measure the dynamical mass of the systems,
\cite{Tacconi02} and \cite{Dasyra06a,Dasyra06b} used near-IR CO and
Si~{\small I} absorption features, tracing the stars in the galaxies,
although \cite{Dasyra06a,Dasyra06b} also used the [Fe~{\small II}]
forbidden emission line, which is a tracer of the warm, ionized
gas. In contrast, \cite{Colina05} used the optical H${\alpha}$
emission line, which also traces the warm, ionized gas. The main
uncertainties associated with these techniques are: (1) the internal
extinction; (2) the light contribution from AGN and/or compact,
nuclear starburst; and (3) the assumption of virialization \cite[for a
detailed discussion see][]{Colina05}. A comparison between our
modelling results and those of the kinematical studies mentioned above
shows that:

\begin{itemize}
\item [-] 
typically the stellar masses are within a factor of $\sim$2 of the
dynamical masses, which is remarkable given the uncertainties in both
spectroscopic and dynamical techniques;

\item [-]
the dynamical masses tend to be larger than the stellar masses in most
cases. However this is not surprising given that the dynamical masses
account for all the mass (including stars, gas and dark matter),
whereas the spectroscopic masses only account for the stars.
\end{itemize}
 
The results presented here reinforce the idea that ULIRGs are
sub-$m_{\ast}$ or $\sim$$m_{\ast}$ systems
\citep{Genzel01,Tacconi02,Colina05,Dasyra06a,Dasyra06b}. They also
provide evidence that the optically visible stellar populations dominate the
stellar masses of the systems; the mass contributions of any stellar 
populations entirely hidden by dust must be relatively minor.

\section{Bolometric luminosities: hidden starburst components}

As well as the total stellar masses, the estimated values for the
bolometric luminosities associated with the stellar populations
detected at optical wavelengths are also presented in Table 4, of
Paper I. For the ES, we find a mean value of L$_{\rm bol}$ = 0.66
$\times$ 10$^{12}$ L$_{\odot}$ and a median of L$_{\rm bol}$ = 0.53
$\times$ 10$^{12}$ L$_{\odot}$. Assuming that most of the optical
light is absorbed and re-processed by dust, it is possible to compare
these values with the mid- to far-IR luminosities of the
soures. Figure \ref{fig:percentage_LIR} presents a histogram showing
the estimated bolometric luminosities of the YSPs detected in the
optical expressed as percentages of the mid- to far-IR luminosities of
the 33 sources in the ES considered here (all but the three Sy1, for
which it was not possible to estimate L$_{\rm bol}$). We find that for
48\% of the objects (16 of 33) the bolometric luminosities of the YSPs
detected at optical wavelengths represent a large fraction ($\gsim$
50\%) of the mid- to far-IR luminosities of the sources.

However, note that 6 of the 16 objects in which L$_{\rm bol}$(YSP)
represents a large fraction of the L$_{\rm IR}$ are spectroscopically
classified as Sy2 galaxies in the optical. Due to potential AGN
contamination, the contribution of the stellar populations to the
optical light, and therefore the bolometric luminosity associated with
them, is less well constrained for these objects than for the other
ULIRGs. Hence, it is possible that the true values of the stellar
bolometric luminosity presented in Paper I are significantly
smaller. Excluding these 6 objects from the group of 16, we find that
in 23 of the remaining 33 objects in the ES (70\%) the bolometric
luminosities of the YSPs represent only a modest fraction of mid- to
far-IR luminosities ($L_{ysp}/L_{ir} <$0.5). On the other hand, there
remain 10 objects in the ES ($\sim$30\%), with no evidence of AGN
activity, for which we find that the YSP luminosities represent a
large proportion of their infrared luminosities. 

At this stage, it is important to mention that there are two major
uncertainties associated to the results presented in this section:
first, uncertainties in the selection of a particular model that is
then applied to the whole galaxy; second, the assumption that all the
stellar light is actually absorbed by, and heats the dust that
radiates at mid- to far-IR wavelengths. In the former case, our model
selection and assumptions are unlikely be substantially in error,
since the stellar masses derived from the same models compare well
with the dynamical masses. Of more concern is the second assumption:
that all the light from the stellar populations heats the dust. In at
least some cases, the reddening of the VYSP in the nuclear aperture is
fairly modest (see paper I for details) and the IYSP reddening is
constrained to be E(B-V) of 0.4 or less. In the extended regions
not covered by the slit (but included in our bolometric luminosity
estimates because of the aperture correction of the flux), the
reddening is likely to be even less. Therefore it is possible that a
substantial fraction of the starlight from the stellar populations
detected in the optical escapes each galaxy, and that the stellar
bolometric luminosities presented in Table 10 in paper I and Figure
\ref{fig:percentage_LIR} in this paper represent over-estimates of the
ability of the starlight to heat the dust.

\begin{figure}
\centering
\hspace{0.0cm}
\psfig{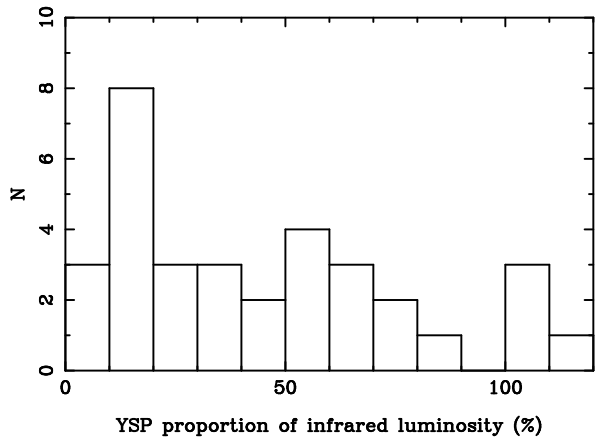}
\caption[Bolometric luminosities of the YSPs expressed as percentages
of the mid- to far-IR luminosities of the objects in the ES]{Histogram
showing the estimated bolometric luminosities of the YSPs detected in
the optical expressed as percentages of the mid- to far-IR
luminosities of the 33 sources in the ES considered here.}
\label{fig:percentage_LIR}
\end{figure}

\section{The nature of the progenitor galaxies}

We discuss in this section the extent to which the stellar populations
detected in the optical have been formed during the merger, and also
the nature of the progenitor galaxies involved in the triggering
events. The aim is to answer two simple, but important,
questions. First, are the YSPs detected in optical observations of
ULIRGS all formed during the merger event? Second, can we use studies
similar to that presented here in order to determine the star
formation histories in major galaxy mergers?

As outlined \cite{Rodriguez-Zaurin08}, the contamination by YSPs
present in the disks of the galaxies prior to the merger is
potentially an important issue. We investigate this by comparing the VYSPs
masses of late-type spirals, likely progenitors of ULIRGs, with those
of the IYSPs in the ULIRGs in our ES. The reason to focus on the VYSPs
in spirals is that, if the star formation is truncated during the
merger, these are the stellar populations that may evolve into the
IYSPs observed today. For the ULIRGs in the ES, we obtain IYSP masses
in the range M$_{\rm IYSP}$ = 0.5 -- 50 $\times$ 10$^{10}$
M$_{\odot}$, with a mean value of 6.0 $\times$ 10$^{10}$ M$_{\odot}$
and a median of 3.4 $\times$ 10$^{10}$ M$_{\odot}$ (see paper
I)\footnote{It is important to emphasize that the upper limit of the
range is found for the radio galaxy IRAS 13451+1232, which clearly
represents an exceptional case among the ULIRGs in the ES}.  In
comparison, \citet{Rodriguez-Zaurin08} estimated that the VYSP of
typical late-type (Sc) spiral galaxies have VYSP masses in the range
M$_{\rm VYSP}$(Sc) = 1.2 -- 8.5 $\times$ 10$^{9}$
M$_{\odot}$. Therefore it is clear that, assuming an equal mass
merger, both of the progenitor spiral galaxies must be at the upper
end of the mass range in order to explain the IYSPs in ULIRGs in terms
of VYSP from the captured disks of the progenitor galaxies; clearly
the ULIRGs with the more massive IYSPs would be difficult to explain
in this way.

Alternatively, we can consider the nature of the progenitor galaxies
that would be required to produce the YSP (IYSP and VYSP) and gas
masses of ULIRGs by gas-rich mergers, with the YSP formed entirely by
merger-induced star formation. The total gas masses M(H$_{2}$ +
H{\small I}) of ULIRGs are typically of order a few times 10$^{10}$
M$_{\odot}$ \citep{Mirabel88,Solomon97,Evans02}.  Adding these gas
masses to the total masses of the IYSP and VYSP (from Table 4 in paper
I) that we are assuming have formed in the mergers, we find a total
mass $M(gas+YSP) > 5\times10^{10}$ M$_{\odot}$ in many of the objects
in our sample. Given that the merger-induced star formation is
unlikely to be more than 50\% efficient, the progenitor galaxies are
together required to have an even larger gas mass. In comparison,
typical late type (Sc) spiral galaxies have total gas masses of
M(H$_{2}$ + H{\small I}) $\sim$ 1.4 $\times$ 10$^{10}$ M$_{\odot}$
\cite[see][for details]{Rodriguez-Zaurin08}. Therefore the progenitor
galaxies are required to be at the upper end of the mass range for
late-type spirals to explain the total YSP and gas masses of the
ULIRGs. Note that, even if we consider only the gas masses, we require
two ``average'' late-type spiral galaxies to merge in order to produce
the typical gas masses measured in ULIRGs, and this is without
accounting for any losses of gas due to outflows and star formation in
the merger.

Finally we note that 33\% of the objects in the ES have VYSPs with
masses M$_{\rm VYSP} \gsim$ 10$^{10}$ M$_{\odot}$ (see paper I). Since
these VYSP must be formed during the merger, and the star formation
efficiency is likely to be $<$50\%, the total amount of gas required
in the progenitor galaxies is much larger; again a merger of two
late-type spirals of average or larger mass would be required to
produce the VYSP alone in these systems.

Overall, regardless of whether the IYSP are formed during the mergers
or merely captured from the progenitor disks, it is clear that the
progenitor galaxies are required to be massive gas-rich spiral
galaxies. Using the compilation of HI masses in \cite{Roberts94}, we
find that, for the cases in which the estimated total progenitor gas
mass is $M(gas+YSP) > 5\times10^{10}$ M$_{\odot}$, {\it both} progenitor
galaxies are required to be in upper 25\% of the gas mass range for
spirals\footnote{Assuming an H$_{2}$/HI mass ratio of 0.7 for spirals
\citep{Young89}.}.

At this stage, it is important to add a caveat about the nature of the
progenitor galaxies. It is commonly accepted that the ULIRG phenomenon
is associated with gas-rich mergers in which both galaxies are spiral
galaxies. This is supported by the mass arguments presented above, as
well as the lack of evidence for a major contribution from an OSP in
most of the ULIRG sample.  However, in the cases of one of the nuclear
apertures in each of the double nucleus systems IRAS 21208-0519 and
IRAS 23327+2913, adequate fits are only obtained with a combination of
stellar populations that includes a dominant contribution of a 12.5
Gyr OSP, plus a VYSP.  Overall, these results suggest that mergers in
which at least one of the progenitors had an early-type morphology can
also trigger the ULIRG phenomenon.

\section{Comparison with merger models}

The range of different morphologies of the objects in the ES, from
widely separated double nuclei systems (e.g. IRAS 14394+5332) to
single nucleus galaxies with no strong signs of tidal structures
(e.g. IRAS 15206-3342), suggests that they are in different
stages of merger events. In this section, we give an overview of how
our modelling results fit in with the merger simulation predictions.

In general terms, the simulations predict two epochs of starburst
activity \citep{Mihos96,Barnes96,Springel05} in major gas-rich
mergers: the first occurring around or just after the first encounter,
and a second more intense
episode towards the end of the merger, $\sim$0.5 -- 1.5
Gyr after the first encounter, when the nuclei are close to coalescing
(within 5$\times$10$^{7}$ Myr of coalescence). However, both the time lag and
the relative intensity of the peaks of starburst activity during the
merger event depend on several factors: the presence of bulges,
feedback effects, gas content and orbital geometry. For example, the
presence of a bulge acts as a stabilizer of the gas against inflows
and the formation of bar structures, allowing stronger starburst
activity towards the end of the merger event
\citep{Mihos96,Barnes96}. On the other hand, AGN feedback effects
(e.g. quasar-driven winds) disrupt the gas surrounding the black hole,
acting against the star formation activity at all stages
\citep{Springel05}.

In Paper I we showed that it is possible to model all of the extracted
spectra with a combination of two stellar populations, plus a
power-law in some cases. Concentrating again on comb III, the
modelling results are, in general terms, consistent with the merger
simulations. For example, it is possible that the IYSPs detected in
most apertures are related to the first enhancement of the starburst
activity, around or just after the time of the first encounter, and we
are now seeing the objects at a later stage.

We emphasize that, although the VYSPs are more
significant in the nuclear regions of the galaxies, this component is
detected at all locations sampled by the extended apertures for the
overwhelming majority of the objects. We also find a trend for the
VYSPs in the nuclear regions to be younger and more reddened than
those of the extended regions. All of these results are consistent
with the merger simulations, which predict that the star formation
activity in the final stages of merger
is concentrated towards the nuclear regions of the merging
galaxies, coinciding with higher concentrations of gas and dust, but
it also occurs within the tidal structures formed during the
interaction. Furthermore, as described in Section 2, when using Comb
I, we find a clear cut-off in YSP ages at 100 Myr in the nuclear
regions. This is in good agreement with the timescale ($\leq$ 100 Myr)
of the induced starburst activity predicted by the theoretical models
for the final stages of a merger, as the nuclei coalesce
\citep{Mihos96,Springel05}.

We note that ongoing star formation activity is also detected for all
objects with relatively wide nuclear separations: IRAS
14394+5332\footnote{Note that in the case of IRAS 14394+5332, the
widest separation system in the sample, the ULIRG phenomenon is not
related with the interaction of the two, widely separated systems, but
with the eastern source, which is itself a merger in its final
stages.} (54 kpc), IRAS 17028+5817 (25 kpc), IRAS 21208-0519 (15 kpc)
and IRAS 23327+2913 (24 kpc). Three of these objects, IRAS 17028+5817,
IRAS 21208-0519 and IRAS 23327+2913, show no signs of AGN
activity. Star formation rates of SFR $\gsim$ 100 M$_{\odot}$
yr$^{-1}$ are required in order to produce the infrared luminosities
(L$_{\rm IR}\geq$ 10$^{12}$ L$_{\odot}$) in these latter objects
\citep{Kennicutt98}. As noted above, such high star formation rates
are only predicted by the merger simulations in two stages during the
merger event:
\begin{itemize}

\item
at, or after, the first encounter, if the parent galaxies have 
insignificant bulges and feedback effects are not important, or
\item
at the end of the merger event, close (within $\sim$50~Myr) to the
coalescence of the nuclei.
\end{itemize}
Estimates of the dynamical timescales for completion of the mergers
ranges from $\geq$50~Myr in the case of IRAS~21208-0519, to
$\geq$80~Myr in the cases of IRAS~17028+5817 and IRAS
23327+2913\footnote{In making these dynamical timescale estimates we
have assumed that the projected distances between the nuclei represent
their true separation, and that the nuclei are moving radially towards
each other with a relative radial velocity of 300 km s$^{-1}$.}
Therefore, if the star formation activity is associated with the final
starburst that occurs around the time of the coalescence of the
nuclei, then this starburst must be relatively long-lived in these
three wide separation systems. Alternatively, the enhanced star
formation activity in these systems may be associated with the
(relatively extended) period of star formation that follows as the
galaxies move apart the first encounter of the two nuclei.  However,
for the latter case, the existing merger simulations do not predict
the large star formation rates ($\gsim$ 100 M$_{\odot}$ yr$^{-1}$) for
the extended period after the first encounter, that would be required
for these systems to be observed as ULIRGs. Most probably, these wide
nuclear separation ULIRGs are systems in which the final starburst has
been triggered relatively early in the period prior to coalescence, as
the nuclei move together.

\section{Evolutionary scenarios}

The presence of evolutionary links between ULIRGs, QSOs/radio galaxies
and elliptical galaxies is commonly accepted. However, the true nature
of the links remain uncertain. While some authors propose the
evolutionary sequence cool ULIRG $\rightarrow$ warm ULIRG $\rightarrow
$ optically selected QSO or radio galaxy (\citetalias{Sanders88a};
\citealt{Canalizo01}), other authors \citep{Genzel01,Tacconi02}
dismiss such an evolutionary sequence based on the location of ULIRGs
in the fundamental plane, similar to that of the intermediate mass
($\sim$ $L_{\ast}$) elliptical galaxies. We will compare here our
study of a sample of 36 ULIRGs with studies of the stellar
populations in samples of QSOs and related objects.

\begin{figure*}
\begin{minipage}{\textwidth}
\begin{tabular}{cc}
\hspace*{1cm}\psfig{file=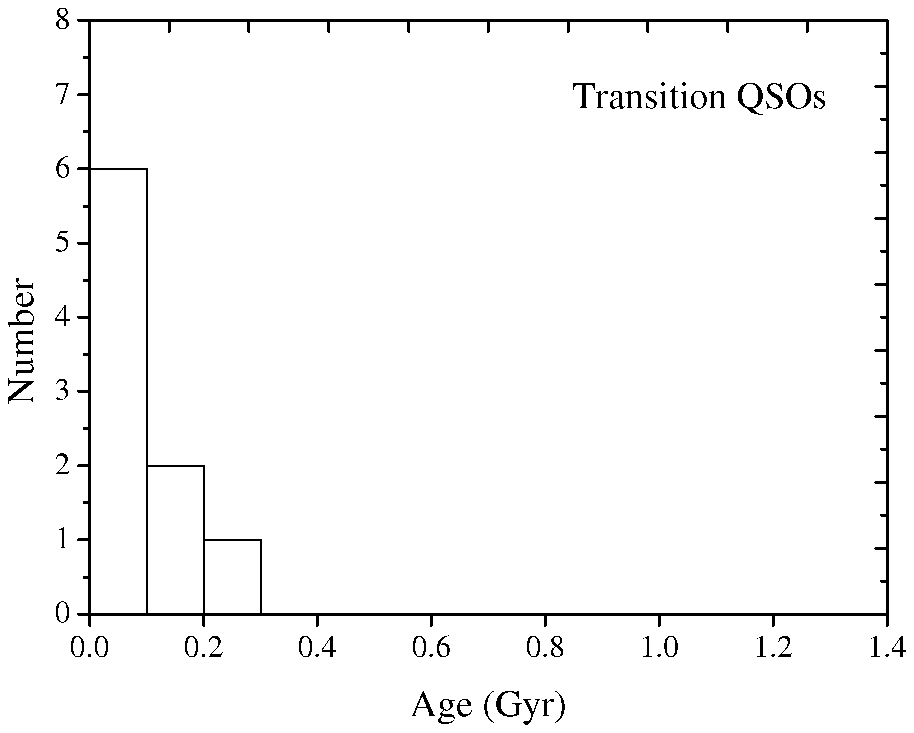,width=7.0cm,angle=0.}&
\psfig{file=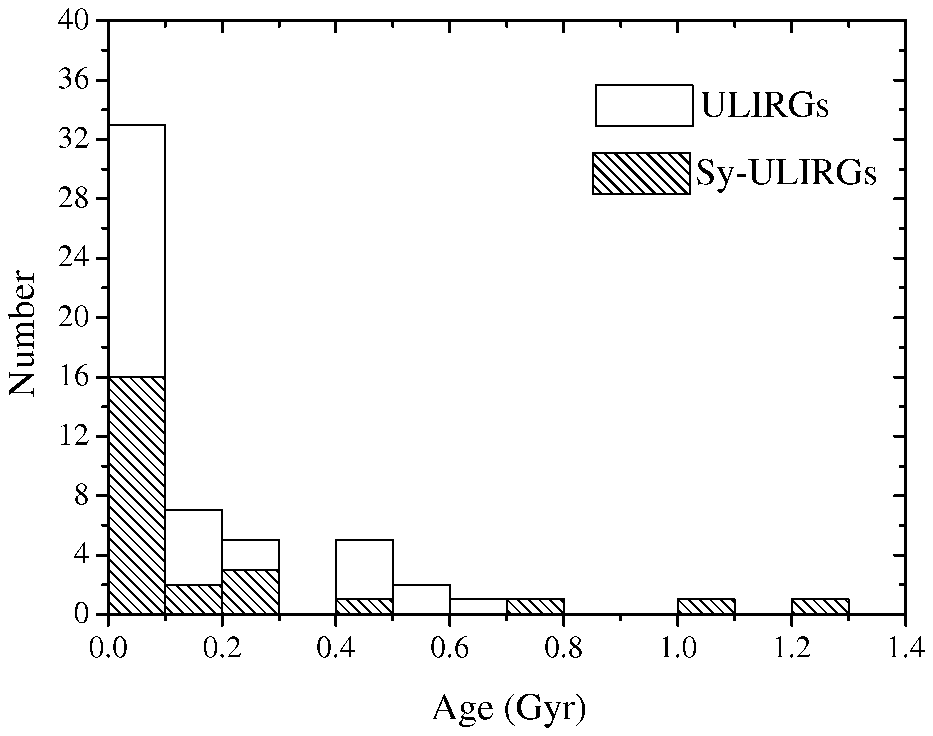,width=7.0cm,angle=0.}
\end{tabular}
\caption[YSPs age distributions of the transition QSOs and the ULIRGs 
in the ES]{The young stellar age distribution for the
\cite{Canalizo01} transition QSOs (left) and the extended regions of 
the ULIRGs in the ES (right), with Seyfert ULIRGs highlighted. These
results are based on modelling combination I. It is clear that the
the transition QSOs, Seyfert ULIRGs, and non-Seyfert ULIRGs overlap in terms 
of their distributions of YSP ages in these diagrams.}
\label{fig:Hist_canalizo}
\end{minipage}
\end{figure*}

\cite{Canalizo00a,Canalizo00b,Canalizo01} carried out a similar study
to that presented here for their sample of 9 ``transition QSOs'' ---
objects that may represent a transitionary stage between ULIRGs and
QSOs. All of their sample objects show clear signs of interaction, and
five of them are classifed as major mergers. Furthermore, all but one
have infrared luminosities L$_{\rm IR} \geq$ 10$^{12}$ L$_{\odot}$
(i.e. they are ULIRGs). In order to study a possible link between
these objects and the general populations of ULIRGs, Canalizo \&
Stockton investigated the stellar populations within the host
galaxies. For that purpose, they modelled the optical spectra
extracted from a series of extended apertures using a combination of a
10 Gyr OSP plus YSPs with varying ages, which is similar to the Comb I
used here. 

Figure \ref{fig:Hist_canalizo} presents histograms comparing the
distribution of the YSP ages derived by
\cite{Canalizo00a,Canalizo00b,Canalizo01} for the transition QSOs,
with those obtained for the ULIRGs in the ES sample using Comb I
(labelled as LW-age in Section 3). For the transition QSOs, we have
used the starburst peak ages \cite[][column 5 of their Table
3]{Canalizo01}, defined as the predominant starburst age found in the
host galaxy \cite[see][for
details]{Canalizo97,Canalizo00a,Canalizo00b}, while for the ULIRGs we
have used the average values of the LW-age, i.e. the same as in Figure
\ref{fig:combI_hist_ES}. Note that, since the nuclear regions could
not be sampled due to contamination by the bright quasar nuclei, the
Canalizo \& Stockton studies concentrate on the extended regions of
the objects. Therefore, we have compared their results with those for
the extended regions of the ULIRGs in the ES. Figure
\ref{fig:Hist_canalizo} also shows the results obtained in this study
for the extended apertures of the Sy-ULIRGs (the ``oldest'' ages in
this figure are obtained for the atypical object PKS1345+12). 

As described in Section 2, we do not find any significant differences
between the ages obtained for the HII- and LINER-ULIRGs and those of
the Sy-ULIRGs in the ES. The same conclusions are reached if we
compare the transition QSOs with the Sy-ULIRGs in the ES sample, or
with the sample of ULIRGs as a whole. If there is an evolutionary link
between ULIRGs as a class and the transition QSOs, one would expect
the latter to have, in general, older YSP ages than ULIRGs, although some
overlap is expected. However, it is clear from the figure that the
ages obtained for the extended apertures in the 
transition QSOs are very similar to those of the
extended apertures of the ULIRGs in the ES. A first interpretation of
this result is that there is no clear evolutionary link between
non-Seyfert ULIRGs and transition QSOs. However, as mentioned in
section 2, it is possible that the timescale for the transition from
ULIRGs to QSOs is smaller than the timescale associated with the ULIRG
phenomenon. 

Therefore, our study does not provide any clear-cut evidence
for an evolutionary link between ULIRGs and the QSOs. Certainly, 
it is unlikely that the AGN in
the Seyfert ULIRGs  in our sample
have been triggered a substantial period ($>$ 100 Myr)
{\it after} the main merger-induced starburst, as appears to
be the case in some radio galaxies \citep{Tadhunter05,Wills08}

\section{Comparison with studies of high-z objects}

Star-forming galaxies have been detected in large numbers at high
redshifts (z $\gsim$ 1). These galaxies are selected on the basis of
their strong UV emission \cite[Lyman-break galaxies -- LBGs: see][for
review]{Giavalisco02}, high sub-millimeter luminosities \cite[sub-mm
galaxies -- SMGs,][]{Smail02,Blain02,Chapman03,Pope05} mid-IR flux
densities \cite[24${\mu}$m {\it Spitzer}-selected
galaxies\footnote{Note that within the so called 24${\mu}$m {\it
Spitzer}-selected galaxies class there will be galaxies classified as
LBGs, SMGs and DRGs. However, not all LBGs and/or DRGs will be detected
at 24${\mu}$m and therefore, we refer to the 
24${\mu}$m {\it Spitzer}-selected objects as a class of galaxies with their own
properties.},][]{Papovich04,Caputi06a,Yan07,Sajina07,Farrah08}, or red
near-IR colours: $(J - K)_{\rm vega} \gsim$ 2.3 \cite[distant red
galaxies -- DRGs,][]{Franx03}. A large fraction show morphological
features suggesting that major mergers are common among such objects
\citep{Giavalisco02,Chapman03,Erb03,Erb04,Conselice05}. In addition,
\cite{Frayer04}, based on deep near-IR imaging, concluded that the
colours, luminosities, morphologies and sizes of the SMGs in their
sample are all consistent with the properties of local
ULIRGs. Furthermore, on the basis of their CO millimeter observations,
\cite{Tacconi06} conclude that SMGs appear to be the ``scaled-up''
version of local ULIRGs. Therefore, it is interesting to compare the
results of this study with those obtained for high-z star forming
galaxies.

\begin{table}
\centering
\begin{tabular}{@{}lcl@{}}
\hline
Star forming & M$_{\rm stellar}$ & references\\
galaxy type         & 10$^{10}$ M$_{\odot}$&      \\
\hline
Lyman-break  &0.2 - 6.0 & Papovich et al. (2001)\\  
Sub-mm       &12 - 25 & Tacconi et al. (2008)\\
Distant Red  &2.9 - 46.0& Papovich et al. (2006)\\  
24${\mu}$m {\it Spitzer}&0.1 - 100 & Caputi et al. (2006)\\  
Local ULIRGs       &0.2 - 50.0& This work\\   
\end{tabular}
\caption{The stellar masses of the different types of high-z star
forming galaxies found in different studies. The stellar masses for
the ULIRGs found in this study are also shown in the table for
comparisson. Note that the upper limit of
50$\times$10$^{10}$M$_{\odot}$ is fouind for the radio galaxy
PKS1345+12, wich represents an exceptional case among the ULIRGs in
our sample.  All results obtained for the high-z star forming
galaxies are based on the modelling of photometric points from the
optical to the mid-IR. }
\label{tab:high-z_masses} 
\end{table}

Table \ref{tab:high-z_masses} shows a compilations of the results
obtained from various studies of the stellar masses of high-z star
forming galaxies. We find that the masses of the local ULIRGs in our
sample are, in general, comparable with or slightly higher than, those
of the LBGs in the \cite{Papovich01} sample. In the cases of the SMGs
and the DRGs, the results found by \citetalias{Tacconi08} and
\citetalias{Papovich06} are consistent with the total stellar mass
estimated in this work for local ULIRGs, and reinforce the idea that
the latter objects are the low redshift analogues of the SMGs and DRGs
found at higher redshifts. On the other hand, we find that the
estimated stellar masses for the ULIRGs in our sample are
significantly smaller (with the exception of the radio galaxy
PKS12345+12) than many of the {\it Spitzer} galaxies. This resuts is
consistent with those of the recent studies of, for example,
\cite{Yan07}, \cite{Sajina07}, \cite{Sajina08}, \cite{Rigby08} and
\cite{Farrah08} among others, suggesting that the {\it Spitzer}
galaxies are either maximum starburst, i.e. scaled-up versions of
local ULIRGs, or indeed, represent a separate class of objects from
local ULIRGs.

Finally, note that the results obtained for the stellar populations of
high-z star forming galaxies are based on fits to relatively few
photometric points, albeit with a wide spectral range. 
The modelling work detailed in Paper I emphasized
the difficulty in obtaining a unique solution when using combinations
of different stellar populations to model spectra, especially on the
basis of fits to the SED alone (i.e. not examining the detailed
absorption features). We have also shown that some results, and
therefore the conclusions based on them, can change significantly with
different modeling assumptions. Furthermore, the SED fits for DRGs in
the \citetalias{Papovich06} sample are relatively poor ($\chi^{2}
\gsim$ 2) in a large fraction ($\gsim$ 50\%) of the sources. Such poor
fits would not be accepted as viable in our study of
ULIRGs. Clearly, particular care must be taken when interpreting the
results obtained on the basis of modelling a few photometric points
using combinations of stellar templates, which is often the
technique used in high-z studies. Studies similar to the one presented
here are vital for increase our undertanding, not only of the
properties of the stellar populations in high-z star forming galaxies,
but also the nature of the links between these objects and the local
ULIRGs.

\section{Summary and conclusions}

In this paper we have presented a detailed analysis of the modelling
results presented for the CS and the ES samples in Paper I. In
addition, we have also discussed these results in a more general
context, comparing them with other studies of ULIRGs. The conclusions
can be summarized as follows.

\begin{itemize}
\item{\bf Age, reddening and percentage contribution}: we find that
the YSPs are more significant in the nuclear than in the extended
regions. The statistical analysis presented here suggests that the
YSPs located in the nuclear regions tend to be younger and more
reddened, although further analysis is required to confirm this
result. All of these results are consistent with the merger
simulations.

\item{\bf Correlations}: we find no evidence for correlations between
the properties of the stellar populations and other properties of
ULIRGs. This can be explained in terms of a selection effect,
i.e. since the objects in the CS and the ES samples are ULIRGs, with
high IR-luminosities and star formation rates, it is likely that we
are observing them at, or after, the first encounter, or at the end of
the merger event, when the nuclei are very close to
coalescing. Therefore, they all have similar properties. A second
possibility to explain the apparent lack of correlations is the
scatter in the properties of the YSPs due to uncertainties in the
modelling technique. A third possible explanation is that other
variables, such as gas content of the parent galaxies and the geometry
of the collision, may have a more important impact on the stellar
properties of ULIRGs than the optical spectral classification,
interaction class or luminosity.

\item{\bf Merger simulations}: in general our modelling results are
consistent with the merger simulations, with the IYSPs in most systems
plausibly associated with the first enhancement of activity following
the first encounter of the nuclei of the merging galaxies, and the
VYSPs likely related to the final enhancement as the nuclei merge
together. 

\item{\bf 100 Myr cut-off}: when using Comb I, we find a clear cut-off
in luminosity weighted YSP ages at 100 Myr for the nuclear apertures,
consistent with the timescale of the merger-induced starburst activity
predicted by the merger simulations in the final stages of the merger
event, as the nuclei coalesce. This result supports the idea that,
despite the varying morphologies, we are observing most of the objects
close to the peak of the star formation activity in the final
stages of the merger event.

\item{\bf The masses of the parent galaxies}: both progenitor galaxies in 
the trigger events must be in the upper 25\% of the mass range of
late-type spirals for most of the ULIRGs in the ES.

\item{\bf The morphologies of the parent galaxies}: although in most 
cases the results are consistent with the idea that the progenitors 
are late-type spirals (albeit unusually massive late-type spirals), in
3 cases (IRAS 13451+1232, IRAS 21208-0519 and IRAS 23327+2913) there
is clear evidence that at least one of the progenitors is an
early-type galaxy. This demonstrates that ULIRGs can potentially be
triggered by mergers between galaxies with a range of galaxy types.

\item{\bf Stellar mass content}: our modelling results suggest that
most ULIRGs are sub-$m_{\ast}$, or $m_{\ast}$ systems ($m_{\ast}$ =
1.4 $\times 10^{11}$ M$_{\odot}$). Such masses support the idea that
these systems will eventually evolve into intermediate mass elliptical
galaxies. The masses obtained from the modelling results are generally
consistent with the dynamical mass estimates (when available) within a
factor of $\sim$ 2.  This result shows that the stellar
populations detected in the optical dominate the stellar masses of the
galaxies. 

\item{\bf Bolometric luminosities:} our results suggest that the
stellar popultions detected at optical wavelengths make a significant
contribution to the mid- to far-IR luminosities of many of
sources. However, due to uncertainties related to the combination of
stellar populations assumed, and to the assumption that all the stellar
emission is absorbed by, and heats, the dust, it is possible that the
values presented in Table 10 in paper I and Figure
\ref{fig:percentage_LIR} in this paper are over-estimates. We conclude
that while the visible stellar populations can make a significant
contribution to the heating of the dust, it is unlikely that they
dominate the heating of the dust in most cases.

\item{\bf The evolution of ULIRGs}: the stellar masses of the ULIRGs
in the ES sample are consistent with the idea that 
ULIRGs evolve into intermediate mass elliptical galaxies.
On the other hand, the data do not provide any clear evidence to
support/refute the evolutionary sequence: cool ULIRG $\rightarrow$
warm ULIRG $\rightarrow$ optically selected QSO, since the stellar
populations of the cold/warm, Sy/non-Sy, QSO/non-QSO ULIRGs are
similar (perhaps the timescale of this transition is shorter than the
timescale of the major merger-induced starburst). However, it remains
possible that some ULIRGs evolve into radio galaxies (although not all
radio galaxies can have this origin).

\item{\bf High-z counterparts}: many of the properties of the ULIRGs 
in our sample, such as morphology, stellar mass content, star
formation histories etc., are consistent with the idea that ULIRGs are
the local counterparts of the  sub-mm galaxies (SMGs) and
distant-red galaxies (DRGs).

\end{itemize}

\section*{Acknowledgments} 
We thank the anonymous referee for useful comments that have helped to
improve this manuscript. JRZ acknowledges financial support from the
STFC in the form of a PhD studentship. JRZ also acknowledges financial
support from the spanish grant ESP2007-65475-C02-01. RGD is supported
by the Spanish Ministerio de Educaci\'on y Ciencia under grant AYA
2007-64712. We also thank support for a joint CSIC-Royal Astronomy
Society bilateral collaboration grant. The William Herschel Telescope
is operated on the island of La Palma by the Isaac Newton Group in the
Spanish Observatorio del Roque de los Muchachos of the Instituto de
Astrofisica de Canarias.

\bibliographystyle{mn2e}
\bibliography{paper_II_reply_part2}

\end{document}